\documentclass[fleqn,usenatbib,useAMS]{styles/mnras}

\usepackage{lmodern}
\usepackage{graphicx}	
\usepackage{amsmath}	
\usepackage{amssymb}	
\usepackage{multicol}   
\usepackage{bm}		    
\usepackage{pdflscape}	
\usepackage{lscape}     
\usepackage{float}      
\usepackage{multirow}   
\usepackage{hyperref}   
\usepackage{todonotes}  

\newcommand{\thecode}{{\sc Sirocco}}
\newcommand{\theoldcode}{{\sc python}}
\newcommand{\sedona}{{\sc sedona}}
\newcommand{\degrees}{$^{\circ}$}
\newcommand{\atomictransition}[2]{#1 \textsc{#2}~}

\defcitealias{dai_unified_2018}{D18}
\defcitealias{Thomsen_2022}{T22}

\usepackage[T1]{fontenc}
\usepackage{ae,aecompl}
\usepackage{newtxtext,newtxmath}

\title[A multi-dimensional view of a unified model]{A multi-dimensional view of a unified model for TDEs}

\author[E. J. Parkinson et al.]
{Edward J. Parkinson$^{1}$\thanks{E-mail: \href{mailto:e.parkinson@soton.ac.uk}{e.parkinson@soton.ac.uk}}, 
Christian Knigge$^{2}$,
Lixin Dai$^{3}$,
Lars Lund Thomsen$^{3}$, \newauthor
James H. Matthews$^{4}$,
Knox S. Long$^{5,~6}$
\\
$^{1}$School of Electronics and Computer Science, University of Southampton, Southampton, SO17 1BJ, UK\\
$^{2}$School of Physics and Astronomy, University of Southampton, Southampton, SO17 1BJ, UK\\
$^{3}$Department of Physics, University of Hong Kong, Pokfulam Road, Hong Kong, China\\
$^{4}$Astrophysics, Department of Physics, University of Oxford, Keble Road, Oxford OX1 3RH, UK\\
$^{5}$Space Telescope Science Institute, 3700 San Martin Drive, Baltimore, MD, 21218, USA\\
$^{6}$Eureka Scientific Inc., 2542 Delmar Avenue, Suite 100, Oakland, CA, 94602-3017, USA\\
}

\date{\today}
\pubyear{2024}

\begin{document}
\label{firstpage}
\pagerange{\pageref{firstpage}--\pageref{lastpage}}
\maketitle


\begin{abstract}
Tidal disruption events (TDEs) can generate non-spherical, relativistic and optically thick outflows. Simulations show that the radiation we observe is reprocessed by these outflows. According to a \textit{unified model} suggested by these simulations, the spectral energy distributions (SEDs) of TDEs depend strongly on viewing angle: low [high] optical-to-X-ray ratios (OXRs) correspond to face-on [edge-on] orientations. Post-processing with radiative transfer codes have simulated the emergent spectra, but have so far been carried out only in a quasi-1-D framework, with three atomic species (H, He and O). Here, we present 2.5-D Monte Carlo radiative transfer simulations which model the emission from a non-spherical outflow, including a more comprehensive set of cosmically abundant species. While the basic trend of OXR increasing with inclination is preserved, the inherently multi-D nature of photon transport through the non-spherical outflow significantly affects the emergent SEDs. Relaxing the quasi-1-D approximation allows photons to preferentially escape in (polar) directions of lower optical depth, resulting in a greater variation of bolometric luminosity as a function of inclination. According to our simulations, inclination alone may not fully explain the large dynamic range of observed TDE OXRs. We also find that including metals, other than O, changes the emergent spectra significantly, resulting in stronger absorption and emission lines in the extreme ultraviolet, as well a greater variation in the OXR as a function of inclination. Whilst our results support previously proposed unified models for TDEs, they also highlight the critical importance of multi-D ionization and radiative transfer.
\end{abstract}



\section{Introduction} \label{sec: intro}

Tidal disruption events (TDEs) occur when a star's self-gravity is overwhelmed by tidal forces during a close encounter with a supermassive black hole (SMBH; \citet{Hills1975a,Martin1988}). Roughly half of the disrupted stellar debris becomes bound to the SMBH, eventually forming a quasi-circular accretion disc that feeds the SMBH \citep{cannizzo_disk_1990,shiokawa_general_2015, hayasaki_circularization_2016,bonnerot_disc_2016, bonnerot_first_2021}. The initial fallback of material and also the early accretion rate of the disc are typically super-Eddington, eventually becoming sub-Eddington as the fallback rate decreases and the mass reservoir empties \citep{strubbe_optical_2009, wu_super-eddington_2018}. Since fallback and accretion convert gravitational potential energy into heat which is radiated away, TDEs are accompanied by a powerful and transient flare visible across the electromagnetic spectrum \citep{saxton_x-ray_2020, van_velzen_optical-ultraviolet_2020, jiang_infrared_2021, alexander_radio_2020, gezari_tidal_2021}. The basic theory of TDEs was developed decades ago \citep[e.g.][]{Hills1975a, young_black_1977, hills_stellar_1978, frank_tidal_1978, Martin1988}, but recent breakthroughs in transient astronomy and numerical simulations have uncovered significant gaps in our understanding. 

If the observed spectral energy distribution (SED) is dominated by thermal emission from the hot inner edge of a quasi-circular accretion disk, we would typically expect it to peak in the (soft) X-ray. However, there are some TDEs with non-thermal X-ray emission thought to be associated with inverse-Comptonization \citep[e.g.][]{Lin2017}. At late-times, the ultraviolet (UV) and optical emission is consistent with thermal emission arising from an accretion disk \citep[e.g.][]{van_velzen_late-time_2019, mummery_spectral_2020, mummery2023, Wen2023}. However, at early-times the characteristic SED temperatures are often significantly lower than expected \citep[e.g.][]{gezari_ultravioletoptical_2012}, and there is a growing population of TDEs whose SEDs peak in the UV or optical band with weak X-ray emission \citep[e.g.][]{2021ApJ...908....4V}. There are at least two ways to account for these observations. First, the luminosity could be generated by powerful shocks that are produced when infalling stellar debris streams collide \citep{dai_soft_2015, 2015ApJ...806..164P,shiokawa_general_2015, ryu_measuring_2020, Steinberg2024}. The other is that emission from a hot accretion disc is reprocessed by an optically thick envelope, such as a mass-loaded outflow, shifting the peak of the SED from the X-ray band to UV/optical wavelengths \citep{strubbe_optical_2009, Metzger2016a, roth_x-ray_2016, roth_what_2018, lu_self-intersection_2019, piro_wind-reprocessed_2020, bonnerot_first_2021, parkinson_reprocessing_2022, Thomsen_2022}.

Reprocessing is likely to be important regardless of which emission mechanism dominates. After all, given the extreme luminosities generated by TDEs at early times, radiation pressure alone would be expected to drive highly mass-loaded and powerful outflows. This expectation has been confirmed by the discovery of broad blue-shifted absorption lines -- a smoking gun signature of outflowing material -- in the ultraviolet (UV) spectra of several TDEs \citep{blagorodnova_broad_2019,hung_discovery_2019, hung_discovery_2020}. These features provide unambiguous evidence for the existence of sub-relativistic outflows, while further possible evidence for outflows comes from blueshifted broad UV emission lines \citep{arcavi_continuum_2014, roth_what_2018, hung_discovery_2019, nicholl_outflow_2020}.
\footnote{{Blueshifted absorption lines represent the "gold standard" when it comes to outflow signatures, because their formation has to take place {\em along the line of sight} to the continuum source. They therefore {\em require} the presence of material moving away from the continuum source. There is no comparable constraint on the spatial relationship between an emission-line-forming region and a continuum source. A blue-shifted emission line must be formed in material that is approaching the observer, but this does not automatically imply that it is moving away from the continuum source.}}
X-ray observations have provided evidence for more highly ionized outflowing gas \citep{miller_flows_2015, kara_ultrafast_2018, Yao_2024}. These outflows could naturally provide the reprocessing medium required to explain redder-than-expected SEDs.

{Three-dimensional (3-D) general-relativistic radiation magnetohydrodynamics (GRRMHD) simulations of TDEs (e.g. \citealt[][(hereafter \citetalias{dai_unified_2018})]{dai_unified_2018}, \citealt{curd_grrmhd_2019, bonnerot_first_2021, Huang_2024}) support this reprocessing picture. In these simulations, the early super-Eddington phase produces powerful mass-loaded outflows and collimated jets.} However, it is not yet possible to predict observable SEDs directly from such simulations, since radiation is treated in a simplified manner. For example, in their modelling of the accretion disc scenario, both \citetalias{dai_unified_2018} and \citet{curd_grrmhd_2019} use the M1 closure relation \citep{levermore_relating_1984}. In this scheme, the radiation field is represented by a single frequency and the ionization state is not calculated. Therefore, to predict the emergent spectra, both \citetalias{dai_unified_2018} and \citet{curd_grrmhd_2019} use radiative transfer codes to post-process their simulations and generate synthetic spectra. The post-processing calculations that have been carried out, to-date, are also subject to significant limitations. For example, \citet{curd_grrmhd_2019} neglect bound-bound opacity and use a simplified treatment for bound-free opacity. \citetalias{dai_unified_2018} include a wider range of opacities, but restrict themselves to a quasi-1-D calculation. Specifically, \citetalias{dai_unified_2018} discretise the outflow into four $\theta$-averaged spherical models to explore the inclination dependence of the observed SED.
 
Despite these limitations, based on the results of their post-processing \citetalias{dai_unified_2018} proposed a unification scenario for X-ray and optically bright TDEs. They propose that the existence of both X-ray bright and optically bright (i.e. X-ray weak) TDEs is explained via an inclination dependence associated with reprocessing in a non-spherical outflow. Polar observers see bare disc emission that has not been reprocessed, thus observing an X-ray bright TDE. Observers at intermediate and equatorial inclinations do not see direct disc emission. Along these sight lines, X-ray (disk) photons are reprocessed, shifting the luminosity away from X-ray wavelengths and toward longer optical wavelengths. The net effect of this reprocessing is enhanced UV/optical emission, but attenuated X-ray emission. In a related study, \citet[][hereafter \citetalias{Thomsen_2022}]{Thomsen_2022} proposed a dynamical unification of UV/optical and X-ray bright TDEs, using a similar approach to \citetalias{dai_unified_2018}. More specifically, \citetalias{Thomsen_2022} simulate and produce post-processed spectra for a TDE accretion disc at multiple accretion rates. In addition to finding the same inclination dependence of the \citetalias{dai_unified_2018} unified model, \citetalias{Thomsen_2022} show that the optically brightness depends also on the accretion rate: larger accretion rates result in denser outflows and greater reprocessing, culminating in an enhanced optical continuum. Thus at later times, as the accretion rate and outflow density decreases, the observed X-ray emission dominates over optical emission for most inclinations.

However, the quasi-spherical radiative transfer calculations carried out by \citetalias{dai_unified_2018} and \citetalias{Thomsen_2022} cannot capture all of the key physics involved in producing the wavelength- and orientation-dependent SED. For example, in a 1-D treatment, photons are forced to diffuse through sight lines of arbitrary optical depth, even though in reality they would preferentially travel and escape along more transparent directions with lower optical depth. Similarly, in 1-D, the dense inner parts of an outflow can effectively shield the outer parts from the ionizing photons produced by the central engine. The same is not true in 2-D, or 3-D, where photons can scatter such shields, leading to much a higher ionization state in the flow \citep[c.f.][]{sim_two-dimensional_2005, Higginbottom2014a, Higginbottom2023syy}. 

Our goal in this work is to carry out detailed 2.5-D ionization and radiative transfer post-processing simulations for the GRRMHD simulation presented by \citetalias{dai_unified_2018}. We use the same snapshot used by \citetalias{dai_unified_2018} for their set of 1-D post-processing calculations, extending the post-processing into 2.5-D. We can therefore more accurately predict the orientation- and wavelength-dependent SEDs and test the reprocessing scenario at the very heart of their unified model. We also check explicitly how 1-D and 2.5-D post-processing simulations differ (both quantitatively and qualitatively), as well as how different assumptions regarding abundances can affect the results. 

\section{Monte Carlo Radiative Transfer Post-processing} \label{sec: method}

Our post-processing calculations were performed using our Monte Carlo radiative transfer and ionisation code for moving media, employing the Sobolev approximation \citep[e.g.][]{Sobolev1957a, rybicki_generalization_1978}. This code, \thecode\footnote{\thecode, previously known as \theoldcode, is an open-source collaborative project available at \hyperlink{https://github.com/sirocco-rt/sirocco}{github.com/sirocco-rt/sirocco}. The name change was made to avoid confusion with the programming language of the same name.}, was initially described by \citet{Long2002}. {Over time, the code has been enhanced and applied to model a variety of systems with biconical outflows, from cataclysmic variables to AGN \citep{Higginbottom2013, Higginbottom2014a, Matthews2015, Matthews2016a}. It has recently been further detailed by \citet{sirocco_release}. Here, we provide only a brief overview and refer readers to \citet{sirocco_release} for further details.}

\subsection{Basic procedure}
    
\thecode\ consists of two separate calculation stages. The first stage calculates the ionization state, level populations and temperature structure of an outflow spatially discretised onto a grid. This is done iteratively by tracking a population of Monte Carlo energy quanta (``photon packets'') which are transported through the grid. Photon packets are randomly generated over a wide wavelength range, sampled from the spectral energy distribution (SED) of the radiation sources included in the simulation. Photon transport occurs in what is referred to as 2.5-D, where photons propagate in 3-D space through a 2-D axisymmetric grid. As the photon packets travel through the grid, they interact with the outflow and update Monte Carlo estimators used to model the radiation field in each cell. The heating effect of photon packets is recorded, and is used to iterate the temperature towards thermal equilibrium, where the amount of heating and cooling in each cell is eventually balanced. 
    
Once the photon packets have been transported through the grid, updated temperature and radiation field estimators are used to recalculate level populations and the ionization state of the outflow. This process is repeated until the simulation has converged. A grid cell is considered to be converged when i) the electron and radiation temperature have stopped changing between iterations to within 5 per cent, and, ii) when the heating and cooling rates are balanced to within 5 per cent. It is usually not necessary, or expected, for all grid cells to converge. Cells with poor photon statistics with noisy Monte Carlo estimators tend not to converge. These cells are usually located near the outer edge of the computational domain, and are generally unimportant to the final result.
    
The second calculation stage produces synthetic spectra for a converged simulation. Additional populations of photon packets are generated, typically over a narrow wavelength range to ensure high signal-to-noise, and flown through the \textit{converged} grid to generate spectra for a selection of defined sight lines.

\subsection{Atomic data} \label{sec: atomic}

Other than for a subset of simulations discussed later in Section \ref{sec: abund-comp}, we adopt \thecode's default solar abundances which are based on \cite{verner_atomic_1995, verner_atomic_1996}. The atomic data we use is based on that outlined by \citet{Long2002}, with improvements described by \citet{Higginbottom2013} and \citet{Matthews2015} {and includes H, He, C, N, O, Ne, Na, Mg, Al, Si, P, S, Ar, Ca and Fe.} We treat H and He with the multi-level ``macro-atom'' formalism of \citet{Lucy2002a, Lucy2003a} and metal lines are treated using a two-level approximation described by \citet{Long2002}. The resulting hybrid scheme is described by \citet{sim_two-dimensional_2005}, \citet{Matthews2015} and \citet{sirocco_release}. 

\subsection{Simulation setup} \label{sec: model_setup}

 \begin{figure}
     \centering
     \includegraphics[scale=0.47]{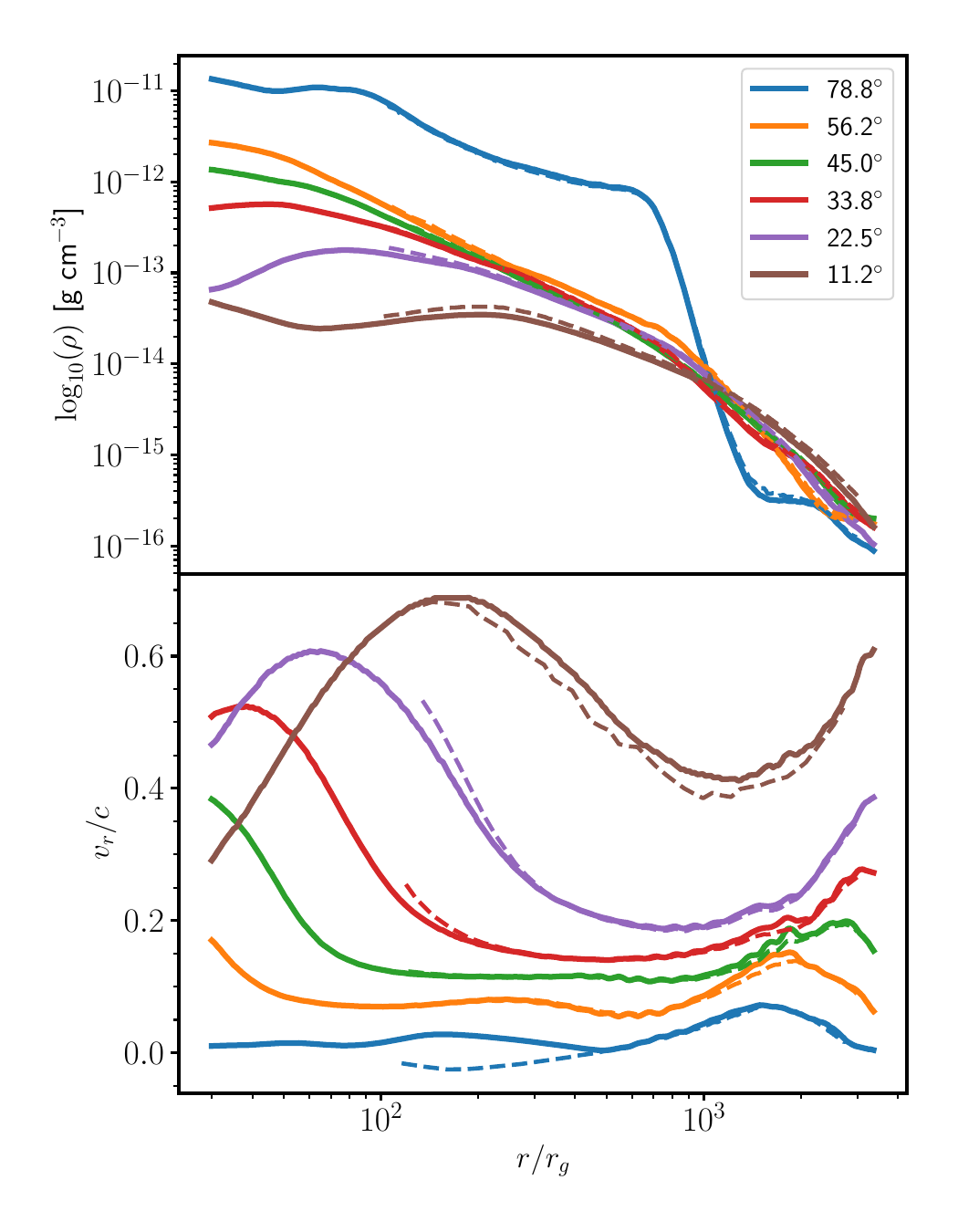}
     \caption{Density and velocity profiles {from \citetalias{dai_unified_2018} and our remapped grid for a selection of sight lines, as a function of $r$. The remapped grid, used in this work, is shown with solid lines, whilst the original is shown with dashed lines.} The profiles shown for the remapped grid are generally in good agreement with the original.}
     \label{fig:rebinned_profiles}
 \end{figure}

\citetalias{dai_unified_2018} presented a 3-D GRRMHD simulation of a super-Eddington accretion disk, such as which may be formed during a TDE. {The initial conditions in this simulation (as well as in those carried out by \citealt{curd_grrmhd_2019} and \citealt{Thomsen_2022}) include a gas torus at large radii that provides the mass reservoir for the accretion disc. The mass and angular momentum of this torus are comparable to those expected for the disrupted star. \color{black} The central object is a SMBH with mass $M_{\mathrm{BH}} \simeq 5 \times 10^6~\mathrm{M_{\odot}}$ and spin parameter $a = 0.8$.}

The 2-D snapshot they analysed and post-processed to develop their unified TDE model was constructed as an azimuthal and time-averaged snapshot over the quasi-steady accretion phase of the simulation. During this phase, the accretion rate was $\dot{M}_{\rm acc} \simeq 15~\dot{M}_{\rm Edd}$, comparable to the peak fallback rate expected during the tidal disruption of a solar-mass star by such a SMBH. In this work, we utilise the same 2-D snapshot that was employed by \citetalias{dai_unified_2018} for their post-processing analysis, extending their 1-D approach into 2-D.

We first have to remap the snapshot onto a different coordinate grid. This is necessary as the GRRMHD simulation by \citetalias{dai_unified_2018} is on a \textit{warped} spherical-polar coordinate grid. \thecode, however, requires the density and velocity structure to be defined onto a structured grid. The new grid we defined splits the snapshot finely into $256 \times 128$ cells, spaced logarithmically in the $r$ direction from $1.3 \text{ to } 8478~r_{g}$, and linearly in the polar $\theta$ direction covering $0^{\circ} < \theta < 90^{\circ}$. This resolution is sufficient to capture even the most finely resolved polar regions in the original warped-grid.  The properties of the snapshot are interpolated onto the new structured grid using \texttt{scrip.interpolate.griddata} \citep{virtanen_scipy_2020}. In \thecode, the density and temperature of each cell are defined at the centre, whereas the cell coordinates and velocities are defined at the inner vertex of each cell. 

{The inner and outer boundaries of our computational domain are set to  $r / r_{g} = 30$ and $r / r_{g} = 4000$, respectively. These choices are guided by the characteristics of the GRRMHD simulation. The inner boundary is placed to avoid the computational challenges of modelling radiative transfer in the very optically thick region and is where the velocity field has transitioned from inflow to outflow in most directions. The outer boundary is chosen conservatively to ensure that only converged regions of the simulation snapshot are included. In the GRRMHD simulation, the accretion disk achieved inflow equilibrium within $r/r_g \simeq 200$, with constant fluxes of mass, energy and angular momentum. Most of the outflowing gas that dominates the reprocessing arises from this steady-state part of the disk. Given the simulation time window from which our snapshot was constructed (between 15,000~$r_g/c$ and 20,000~$r_g/c$), as well as the characteristic outflow speeds (see Figure~\ref{fig: model_properties}), all but the outermost regions near the equatorial plane should also have reached outflow equilibrium. Since edge-on viewing angles may cut through non-equilibrium parts of the simulation -- yet are by far the most optically thick (see Figure~\ref{fig: optical_depth_spectra}) -- the predicted spectra for these viewing angles are not as reliable.}

We have removed the region dominated (and evacuated) by the relativistic jet launched in their simulation, corresponding to polar angles $\theta < 5^{\circ}$. This is done by setting $\rho = 0$ in this region. Figure \ref{fig:rebinned_profiles} shows the density and radial velocity of our remapped grid for six different inclinations. The density and velocity measured on the native (warped) grid are shown as dashed lines, while the same parameters after remapping are shown as solid lines. The level of agreement between original and remapped values is adequate throughout the computational domain. 

The snapshot constructed by \citetalias{dai_unified_2018} for their post-processing analysis does not include information about the rotational velocity in the flow $v_{\phi}$\footnote{{The GRRMHD simulation itself did include this information. However, it was discarded in constructing the snapshot for post-processing and is no longer available.}}. This is an important limitation for {\em all} post-processing efforts; those carried out by \citetalias{dai_unified_2018} and those presented here. While the neglect of rotation should not impact the global reprocessing picture being tested, it will affect the detailed radiative transfer and, in particular, the appearance of spectral lines. We have included only one quadrant of the simulation ($0^{\circ} < \theta < 90^{\circ}$) because \thecode\ is designed to model outflows that are symmetric about the equatorial plane and the rotation axis. We do not expect this to significantly affect our results, since both quadrants in the original snapshot share similar density and velocity structures \citepalias[see Figure 2 in][]{dai_unified_2018}.
 
Prior to performing the spectral synthesis in this work, we verified that our remapped grid is sufficiently similar by re-constructing the 1-D models by \citetalias{dai_unified_2018} and re-producing their post-processing results. We found the spectra generated with \thecode\  are in good agreement with \citetalias{dai_unified_2018}. This is described in further detail in Appendix \ref{sec: python_verification}.

\subsection{Input radiation field}  \label{sec: radiation_sources}

Following \citetalias{dai_unified_2018}, we model the input radiation field by including an isotropically emitting central point source. Photons are injected into the simulation domain from the centre, with a frequency and luminosity uniformly sampled from a blackbody SED characterised by temperature $T_{\mathrm{BB}} \simeq 2.2 \times 10^{5}$ K, which is the radiation temperature of the GRRMHD simulation at our selected inner boundary of $r / r_{g} = 30$. The luminosity of our SED is $L_{\mathrm{BB}} \simeq 12~L_{\rm Edd}$ following from the accretion rate found in \citetalias{dai_unified_2018}'s simulation. In reality, the emission from the inner accretion region is likely to be anisotropic with an SED much more complex than a simple blackbody. However, the structure, evolution and stability of such disks and the radiation fields they produce are active areas of research and still highly uncertain \citep[e.g.][]{hirose_radiation-dominated_2009, jiang_thermal_2013, blaes14, Shen2014}. The use of a simple input radiation field also makes it easier to i) isolate and understand the impact of reprocessing and ii) compare to previous work and understand the multi-D effects.

\subsection{Reprocessing mechanisms} \label{sec: ch5_reprocessing_intro}

{We assume the outflow is in radiative equilibrium, except for adiabatic cooling. Energy absorbed by the outflow is reprocessed via two main mechanisms: (i) re-radiation (i.e. emission) and (ii) conversion to $PdV$ work. We will often refer to re-radiation as "atomic reprocessing", since it describes reprocessing via atomic processes, such as radiative recombination, free-free emission, line emission or via Compton scattering. All of these processes are also included in the heating and cooling balance of the plasma. The other mechanism is the conversion of radiative energy into $PdV$ work. This is referred to as ``adiabatic reprocessing'' by \citet{roth_what_2018}, although we prefer to describe it as "bulk scatter reprocessing." }Both terms refer to the redshifting of the SED as photons undergo successive electron scatterings in a diverging, optically thick outflow \citep[see, e.g.,][]{titarchuk_downscattering_2005, laurent_effects_2007, roth_x-ray_2016, roth_what_2018}. Bulk scatter reprocessing depends on the divergence and velocity of the outflow, and is the effect of the Doppler shift when transforming between the co-moving frame of the outflow and the observer frame. {Generally speaking, in a diverging flow, where $\nabla \cdot \mathbfit{v} > 0$, photons are successively redshifted each time they scatter in the outflow. In a converging flow (e.g. inflow) where $\nabla \cdot \mathbfit{v} < 0$, the opposite occurs and photons are blueshifted.}

Broadly speaking, \citetalias{dai_unified_2018} find that bulk scatter reprocessing is important throughout the entire outflow of their TDE simulations, as photons are trapped by the high optical depths along virtually all directions. Along the mid-plane, densities are higher and ionization states are lower, so the dominant reprocessing mechanism for equatorial (edge-on) sight lines is reprocessing from atomic processes. Photons in this region are absorbed by, for example, the photon-ionization of H and He. This absorbed energy is re-emitted at longer wavelengths via recombination, free-free and/or line emission \citep{roth_x-ray_2016, roth_what_2018, dai_unified_2018}. The thermal state of electrons and Compton scattering also modify the spectrum, by up- and down-scattering photons. 

\section{Results} \label{sec: results}

\subsection{Physical properties} \label{sec: physical_properties}

\begin{figure*}
 \centering
 \includegraphics[scale=0.55]{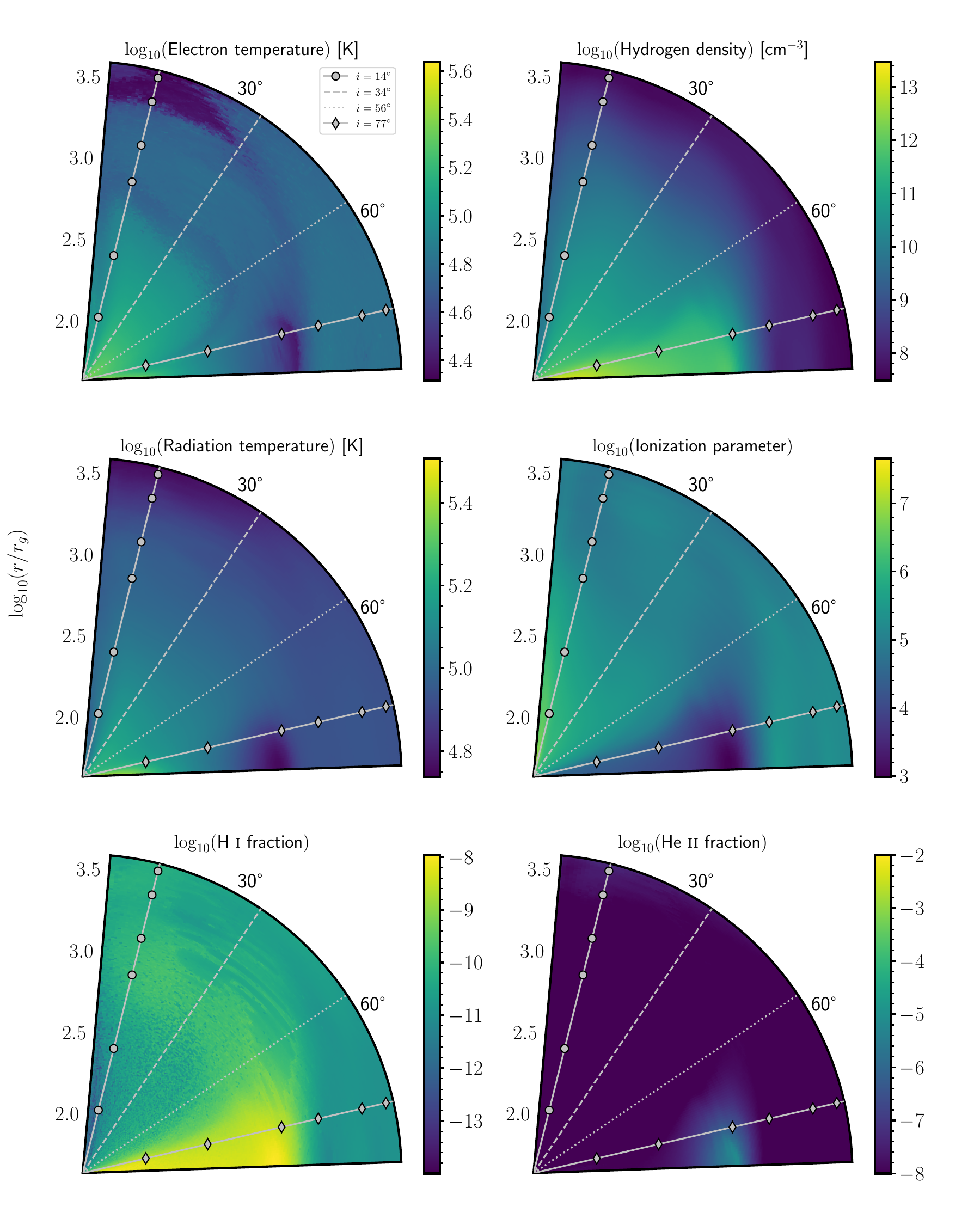}
 \caption{Colour plots for a selection of physical parameters from the output of the 2.5-D radiative transfer simulation, constrained to $5^{\circ} < \theta < 88^{\circ}$. {The grey lines correspond to sight lines, which are labelled in the legend in the top left panel. The markers for the $i=14$\degrees~and $i=77$\degrees~sight lines correspond to the radial sample points in Figure \ref{fig: cell_seds}. All quantities are calculated from our radiative transfer simulation, other than the H density, which was calculated in \citetalias{dai_unified_2018}'s GRRMHD simulation.} \textit{Top left:} electron temperature. \textit{Top right:} H density.  \textit{Middle left:} radiation temperature. \textit{Middle right:} ionization parameter. \textit{Bottom left:} H \textsc{i}  ion  fraction. \textit{Bottom right:} He \textsc{ii} ion fraction.}
 \label{fig: model_properties}
\end{figure*}

A selection of physical properties of our simulation is shown in Figure \ref{fig: model_properties}. However, not represented in the figure is the convergence of the outflow. Approximately $70\%$ of the cells are fully converged, with the remaining struggling to converge in one of the criteria.  The cells struggling to converge are located at the outer edge of the outflow, where the density is low and photon statistics are poorer. Due to the low density and high ionization state, these cells contribute very little to the generated spectra.

The electron temperature of the outflow is shown in the top left panel. It is the largest at the launching point of the outflow, reaching temperatures of $\log_{10}(T_{e}) \gtrsim 5.4$ K. The electron and radiation temperature (middle left) are roughly equal here. As the radius increases, both the electron and radiation temperature decrease. At polar and equatorial angles at $\log_{10}(r / r_{g}) \sim 3.3$, the electron temperature suddenly drops to $\log_{10}(T_{e}) \sim 4.4$. This cooler material is a result of line cooling by collisional excitation, producing a family of EUV emission lines. {Line cooling is able to dominate in these regions because the radiation field has become sufficiently soft, and the ionization state sufficiently low. Under these conditions, bound-bound transitions can provide an efficient emission, and hence cooling, mechanism.}

The H density (top right) tracks the matter density. The most dense region corresponds to the accretion disc region along the mid-plane. At the inner boundary, the H density reaches $\log_{10}(n_{\rm H}) \sim 13$ cm$^{-3}$ and decreases to $\log_{10}(n_{\rm H}) \sim 12$ cm$^{-3}$ by $\log_{10}(r / r_{g}) \sim 2.7$ before dropping further past $\log_{10}(r / r_{g}) \sim 2.7$. For intermediate regions, the density is not as large and is roughly $\log_{10}(n_{\rm H}) \sim 11$ cm$^{-3}$ in the inner region, decreasing to $\log_{10}(n_{\rm H}) \sim 7$ cm$^{-3}$ by the outer boundary. Along polar directions ($i < 10$\degrees), the gas is more dilute and is typically an order of magnitude, or two, less dense than along intermediate inclinations. 

Throughout most of the wind, H and He are almost entirely ionized. The edge of the disc  atmosphere, which has been shielded, houses a small enhanced population of both neutral and singly ionized He and neutral H (bottom right and bottom left panels respectively). This part of the outflow is illuminated by an SED which has been reprocessed and attenuated, resulting in a lower ionization state. The level of reprocessing can be traced using the radiation temperature (middle left panel), which represents the mean photon frequency of a blackbody radiation field, defined as,
\begin{equation}
 T_{r} = \frac{h \bar{\nu}}{3.832~k_{B}},
\end{equation}
where $\bar{\nu}$ is the average photon frequency (in a cell), $h$ is Planck's constant and $k_{B}$ Boltzmann's constant. At the same position where H and He are in a lower ionization state, the radiation temperature is lower. This shows that this region is illuminated by a softer SED, since the average photon frequency is lower. The radiation temperature is also lower in the cooler regions, suggesting there is a softer SED here as well. To quantify the ionization state, it is possible to use the ionization parameter, $U_{\text{H}}$, where,
\begin{equation}
 U_{\text{H}} = \frac{4\pi}{n_{\text{H}}~c} \int_{13.6 \frac{\text{eV}}{h}}^{\infty}
 \frac{J_{\nu}}{h\nu}~d\nu,
 \label{eq: ion_param}
\end{equation}
and where $\nu$ denotes frequency, $n_{\text{H}}$ is the number density of H, $c$ is the speed of light, $h$ is Planck's constant and $J_{\nu}$ is the monochromatic mean intensity. The ionization parameter measures the ratio of the number density of H ionizing photons to the number density of H, making $U_{\text{H}}$ a useful predictor of the global ionization state. However, $U_{\text{H}}$ has no knowledge of the SED shape, meaning it is a poor indicator for the ionization state of other ionic species. In the outflow, the ionization parameter is fairly uniform throughout, with $\log_{10}(U_{\rm H}) \sim 5.5$ meaning H is ionised. However, the ionization parameter is substantially lower in the disc along the mid-plane at $\log_{10}(r / r_{g}) \lesssim 2.5$. This region is illuminated by a reprocessed SED, meaning a reduced number of H ionizing photons. However because $\log_{10}(U_{\rm H}) \sim 3$, there is actually a very small population of neutral H as shown in the bottom left panel, where the disc atmosphere has a larger ion fraction for \atomictransition{H}{i}. Along low inclination angles, at the inner boundary, the ionization parameter is large. This comes from high-energy photons being able to escape along this path of low optical depth and ionizing the outflow. However, by $\log_{10}(r / r_{g}) \sim 2.7$, the ionization parameter has been moderated via radiative and bulk scatter reprocessing.

\subsection{Optical depth} \label{sec: optical_depth}

\begin{figure}
 \centering
 \includegraphics[width=\linewidth]{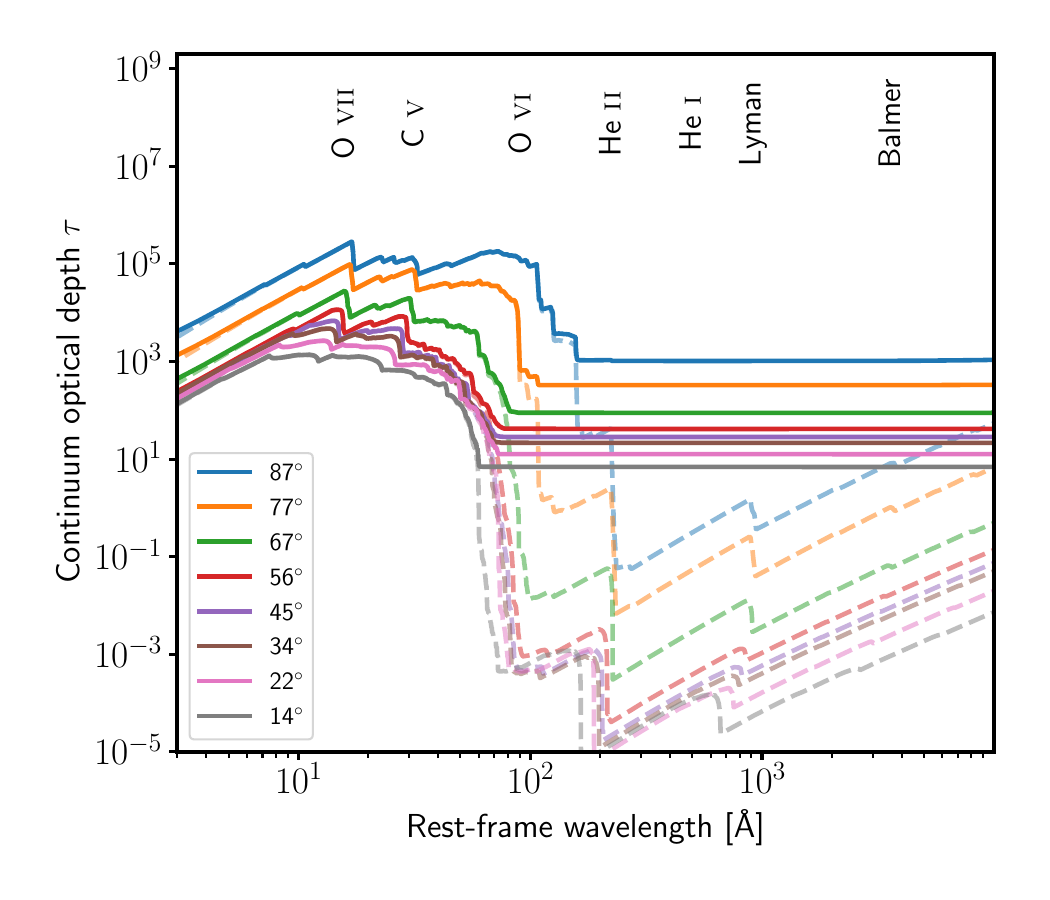}
 \caption{The continuum optical depth, integrated from emission to escape, as a function of frequency for various sight lines with electron scattering included (solid lines) and without electron scattering (dashed lines). Along the base of the wind, there is significant opacity due to the photo-ionization of \atomictransition{He}{ii}, \atomictransition{O}{vi} and \atomictransition{O}{vii}.}
 \label{fig: optical_depth_spectra}
\end{figure}

\begin{figure*}
 \centering
 \includegraphics[scale=0.55]{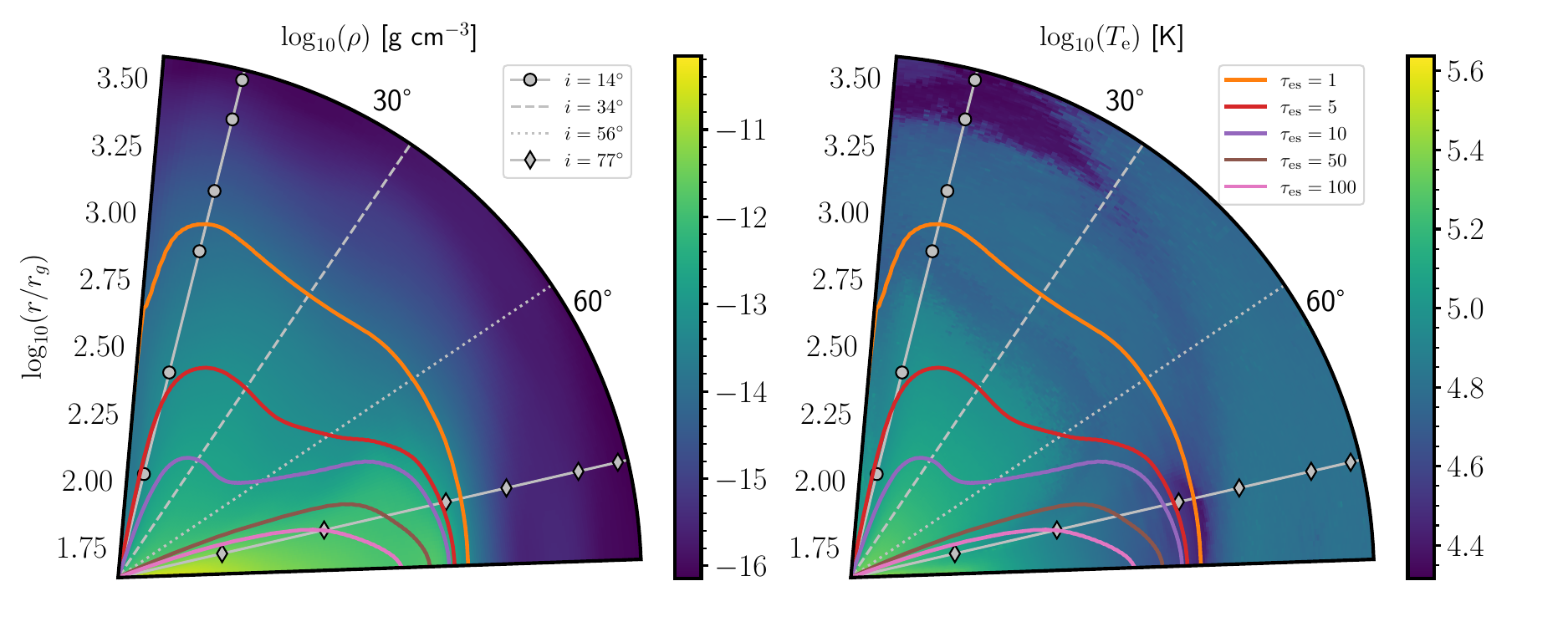}
 \caption{Colour plots showing the the mass density (left) and electron temperature (right) of the outflow and solid lines showing the outline of multiple electron scattering optical depth surfaces for values of $\tau_{\rm es}$ labelled in the legend. {The grey lines in each panel correspond to sight lines, which are labelled in the legend in the left panel. The markers for the $i=14$\degrees~and $i=77$\degrees~sight lines correspond to the radial sample points in Figure \ref{fig: cell_seds}.} The outflow is optically thick to electron scattering ($\tau_{\rm es} > 1$) out to a radius of $1000~r_{g}$, with the highest electron scattering in the disc atmosphere which is traced by the $\tau_{\rm es} = 50$ surface.}
 \label{fig: optical_depth_surfaces}
\end{figure*}

The continuum optical depth for multiple sight lines is shown in Figure \ref{fig: optical_depth_spectra}, with (solid lines) and without (dashed lines) electron scattering included. For the majority of the wavelength range considered, the scattering optical depth dominates over any other source of opacity, such as photo-ionization or free-free absorption. For polar sight lines, $\tau_{\rm es} \lesssim 10$. For higher inclination observers, $\tau_{\rm es} \sim 50$ but this reaches closer to $\tau_{\rm es} \sim 1000$ along the mid-plane. {There is also significant opacity from photon-ionization of singly ionized He and other metals such as O and C at short wavelengths. This suggests that, close to the mid-plane, atomic reprocessing (due to photo-absorption) will be efficient for high-energy photons.}

Even though the scattering optical depth dominates throughout most of the outflow, by removing its contribution to the opacity it is revealed that there still is significant opacity from photo-ionization. Close to the mid-plane, photo-ionization of the H Lyman and Balmer edges, as well as \atomictransition{He}{i} are large with $\tau \sim 10$ but are not as important than, for example, \atomictransition{He}{ii} or \atomictransition{C}{v} when it comes to reprocessing the emission. 

Several electron scattering ``photospheres'' (strictly speaking, constant optical depth surfaces) are shown in Figure \ref{fig: optical_depth_surfaces}, for radially integrated inward optical depths of $\tau_{\rm es} = 1,~5,~10,~50 \text{ and, } 100$. The $\tau_{\rm es} = 1$ photosphere is located at $\log_{10}(r / r_{g}) \sim 3$, but moves in closer to the central source for low inclinations close to the polar jet region. For larger values of $\tau_{\rm es}$, the surfaces are located nearly entirely within the accretion disc atmosphere, with $\tau_{\rm es} = 50$ almost tracing out this boundary. Along lower inclination angles ($\theta < 15^{\circ}$), the surfaces are located further in, suggesting that photons travelling along these sight lines travel more freely and will be reprocessed less by the outflow. This should mean that polar observers are far more likely to see \textit{bare} un-reprocessed SED emission. However, there is nothing to stop reprocessed photons from other parts of the outflow from escaping along these lower optical depth inclinations.

\subsection{Synthetic spectra} \label{sec: synthetic_spectra}

\begin{figure*}
 \centering
 \includegraphics[scale=0.55]{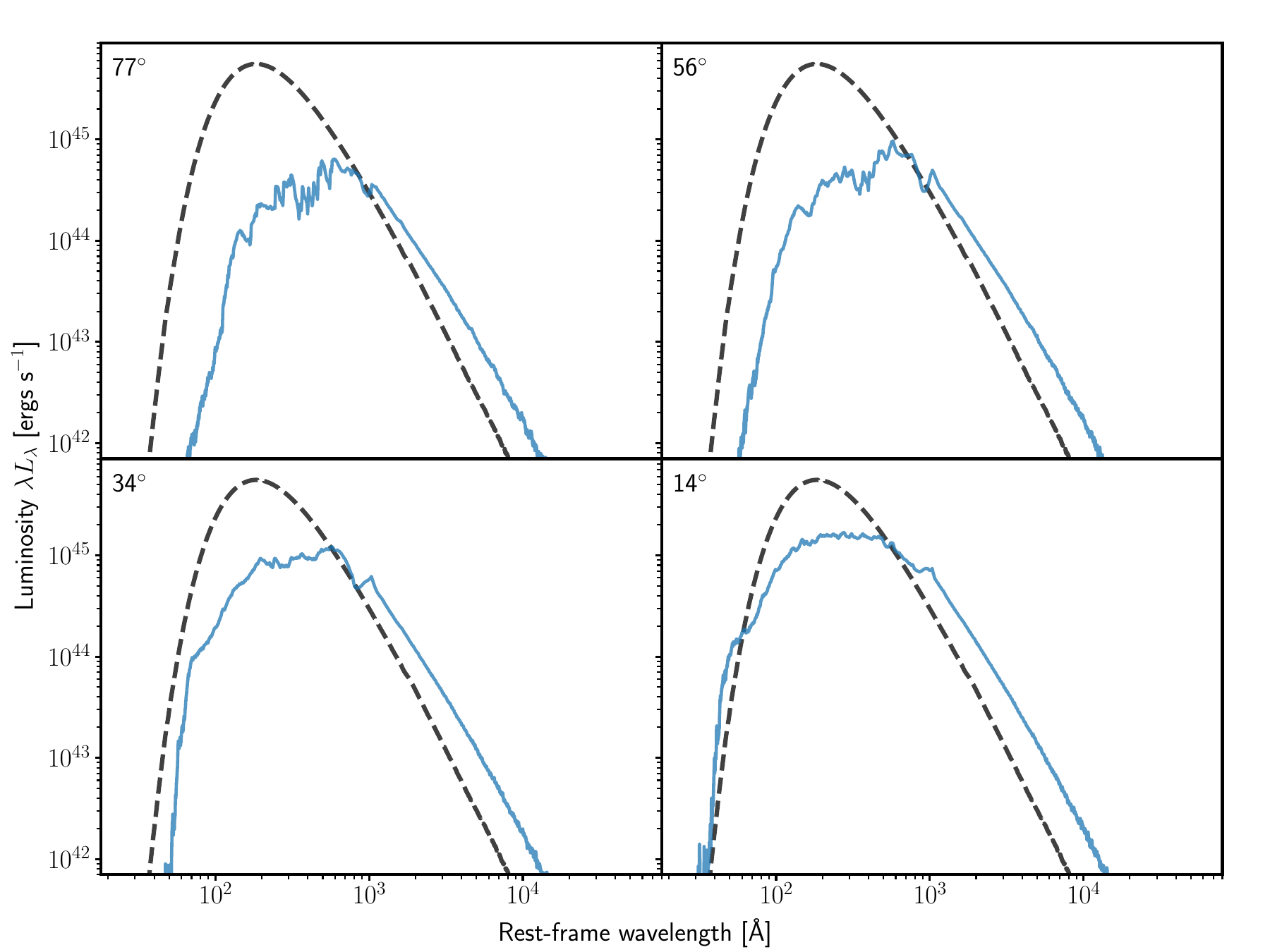}
 \caption{Rest-frame synthetic spectra for four sight lines. Each panel shows a different inclination spectrum, which has been labelled in the top left, as well as the input SED  shown by a dashed black line. The spectra have been smoothed. Each spectrum has been reprocessed, resulting in the emergent spectra taken the form of a stretched and red-shifted blackbody SED, with an enhanced optical continua.}
 \label{fig: model_sed}
\end{figure*}

In Figure \ref{fig: model_sed} we show synthetic spectra generated along four sight lines (labelled in each panel), compared to the input SED. The synthetic spectra take the shape of a modified version of the input spectrum, that is to say that they take the form of a reprocessed (multi-colour) blackbody red-shifted to longer wavelengths, with an enhanced optical continuum relative to the input. This trend exists for all four sight lines, however the \textit{amount} of and \textit{dominant} mechanism of reprocessing, which is reflected in the details of the spectra, changes with inclination. The luminosity escaping is roughly an order of magnitude lower than what is injected into the outflow: whilst $L_{\mathrm{BB}} = 12 L_{\mathrm{Edd}}$ was injected, only $L \sim 2 L_{\mathrm{Edd}}$ escapes. Most of the luminosity is lost due to photons back-scattering into the absorbing inner boundary, which are removed from the simulation. {Whilst this is far from ideal, we are not attempting to model the radiation source and are using, instead, an idealised single-temperature blackbody. The aim of our work is to correctly model the matter and radiation field at radii larger than the inner boundary. Our approach here is consistent with the radiation field in the original GRMMHD snapshot, which has a low radiative efficiency, and is also consistent with \citetalias{dai_unified_2018} and \citetalias{Thomsen_2022}.}

Ignoring the details of the spectra, the trend of the optical-to-X-ray (OXR) emission increasing with inclination found by \citetalias{dai_unified_2018} is also found in our synthetic spectra. The lowest inclination (i.e., closest to face-on) spectrum appears the least reprocessed and is most similar to the input. However, even in this case, bulk scatter reprocessing has increased the \textit{spread} of photon energies, which has stretched the spectrum across a wider wavelength range. There is also a high energy ``shoulder'' in the lowest inclination spectrum, {created by photons scattering in regions where the velocity field is inflowing.} Whilst the amount of atomic reprocessing is low in the outflow when looking at low inclination angles, there is still an enhanced optical continuum. This is both a result of bulk scatter reprocessing shifting the spectrum, and also a consequence of atomically reprocessed photons coming from the optically thick parts of the wind. These photons escape, and are beamed, along sight lines of low optical depth, something which \textit{cannot} happen in a 1-D geometry. So whilst a low inclination observer is more likely to see bare SED emission, they also see reprocessed photons which are beamed along columns of low optical depth. This multi-D transport effect results in the optical continuum being fairly insensitive to inclination (this is shown clearer later in Figure \ref{fig: oxr_sed}), as reprocessed photons escape in all directions, no matter where they come from.

The highest inclination spectra, which look \textit{into} the optically thick region of the outflow, show signatures of photons being reprocessed by atomic interactions. The high-energy tail of the spectra is significantly absorbed, primarily from photo-ionization. Photons which push and scatter their way along the mid-plane of the outflow become trapped by the high optical depth (see Figure \ref{fig: optical_depth_surfaces}), so are reprocessed multiple times. Along these high inclination sight lines, the proportion of X-ray photons contributing to the observer spectrum is small, as most of the energy escapes at EUV and longer wavelengths.

The spectra also contain strong EUV emission and absorption features, particularly at the two lowest inclinations. These lines and the EUV continuum are difficult to observe, as EUV photons are easily extinguished by interstellar absorption in the host galaxy and/or the Milky Way. The fact that a large amount of the (reprocessed) emission is in the EUV could help solve the TDE missing energy problem and merits further investigation \citep[e.g.][]{2015ApJ...806..164P, Lu_2018, Thomsen_2022}. 

\section{Discussion} \label{sec: discussion}

\subsection{The optical-to-X-ray ratio as a function of inclination} \label{sec: optical_xray_ratio}

\begin{figure}
 \centering
 \includegraphics[scale=0.48]{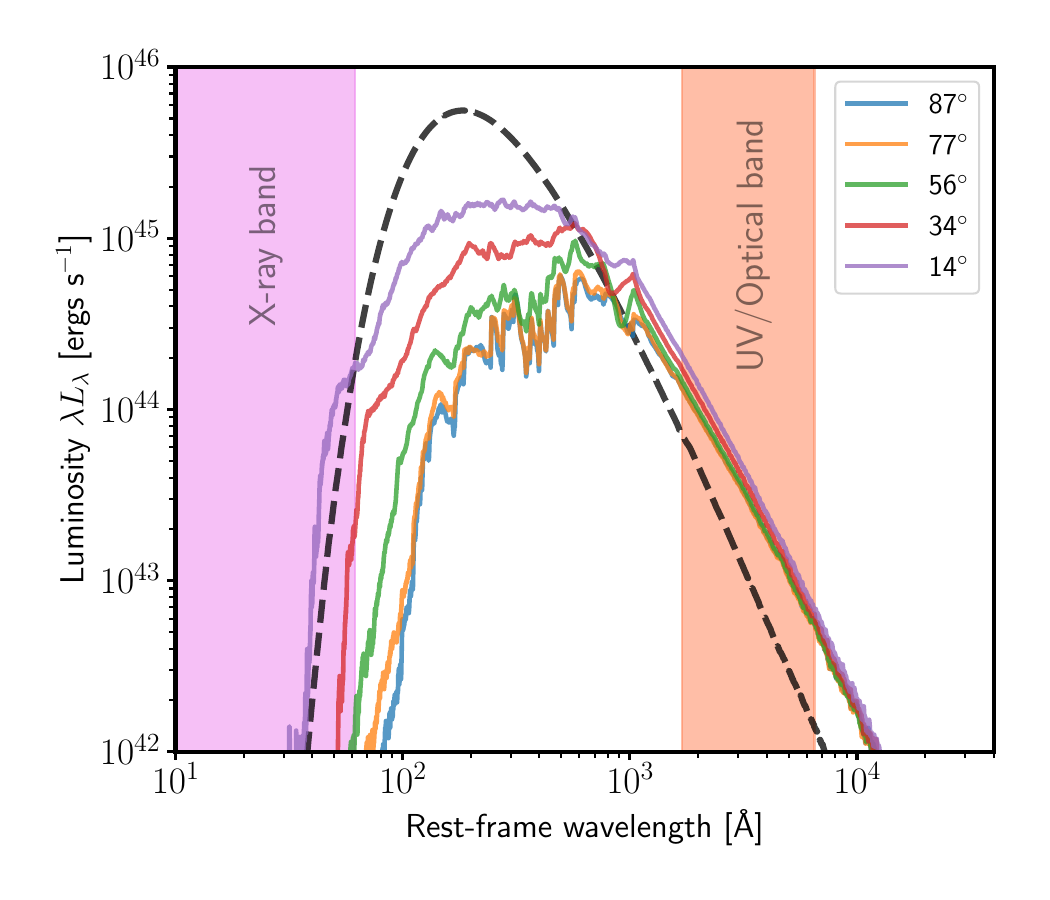}
 \caption{Rest frame synthetic spectra for five sight lines, labelled in the legend). The input SED is shown in black. Shown by the shaded purple and coral areas are the X-ray and optical bands. The X-ray band has been defined to include photons with energies 0.2 keV and above, and the optical waveband is defined between $1700$ - $6500$ \AA. The OXR ratio increases with inclination angle. At high inclinations, observers see a bright optical spectrum with little accompanying X-ray emission. This result is consistent with the unified model proposed by \citetalias{dai_unified_2018}.}
 \label{fig: oxr_sed}
\end{figure}

\begin{figure}
 \centering
 \includegraphics[scale=0.48]{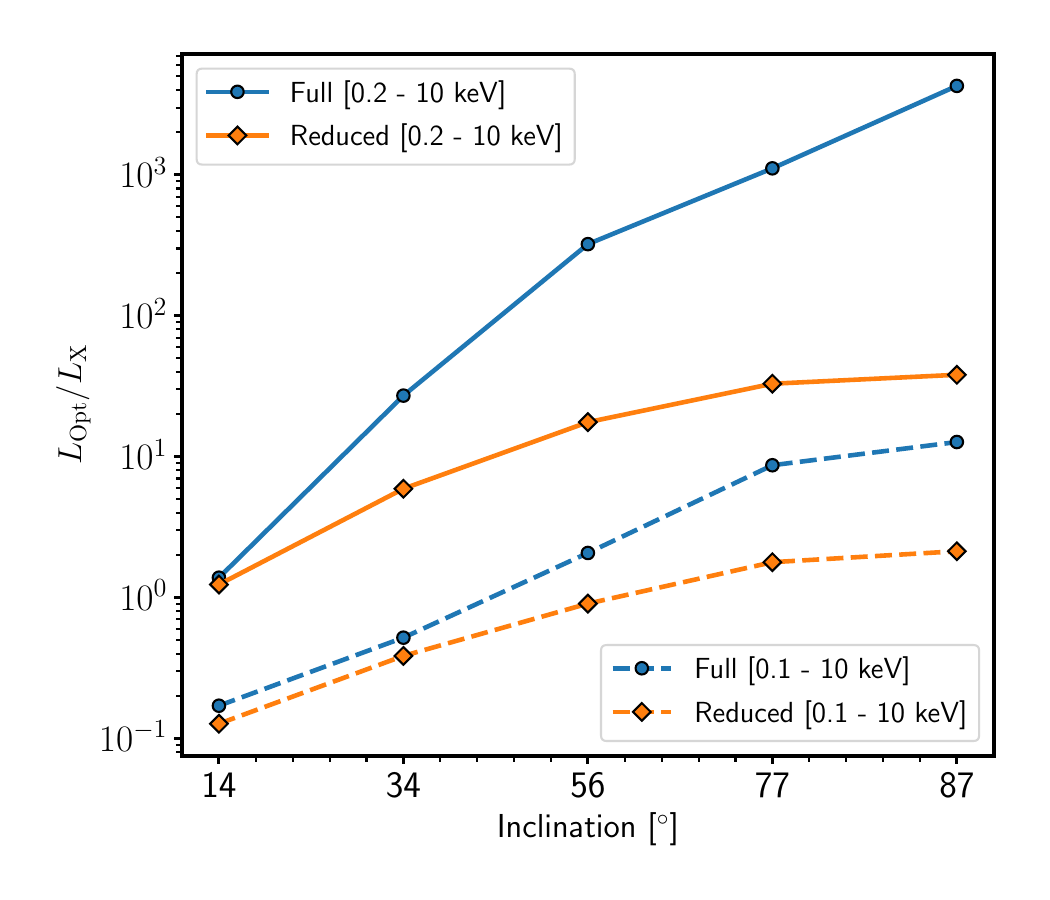}
 \caption{The integrated OXR luminosity ratios as a function of inclination angle. The X-ray band has been defined in two ways: one band include photons with energies 0.2 keV and above, and another includes photons with energies 0.1 keV and above. The optical wavelength is defined between $1700$ - $6500$ \AA. Two sets of results are shown: ``Full'' refers to the simulation shown in Figure \ref{fig: oxr_sed} and ``reduced'' refers to a simulation with only H, He an O (this is discussed in more detail in Section \ref{sec: abund-comp}). The UV/optical luminosity of the spectra increases with inclination, whilst the high-energy luminosity decreases. The ratio depends sensitively on how the bands are defined, as well as the input SED.}
 \label{fig: oxr}
\end{figure}

\begin{figure}
    \centering
    \includegraphics[scale=0.48]{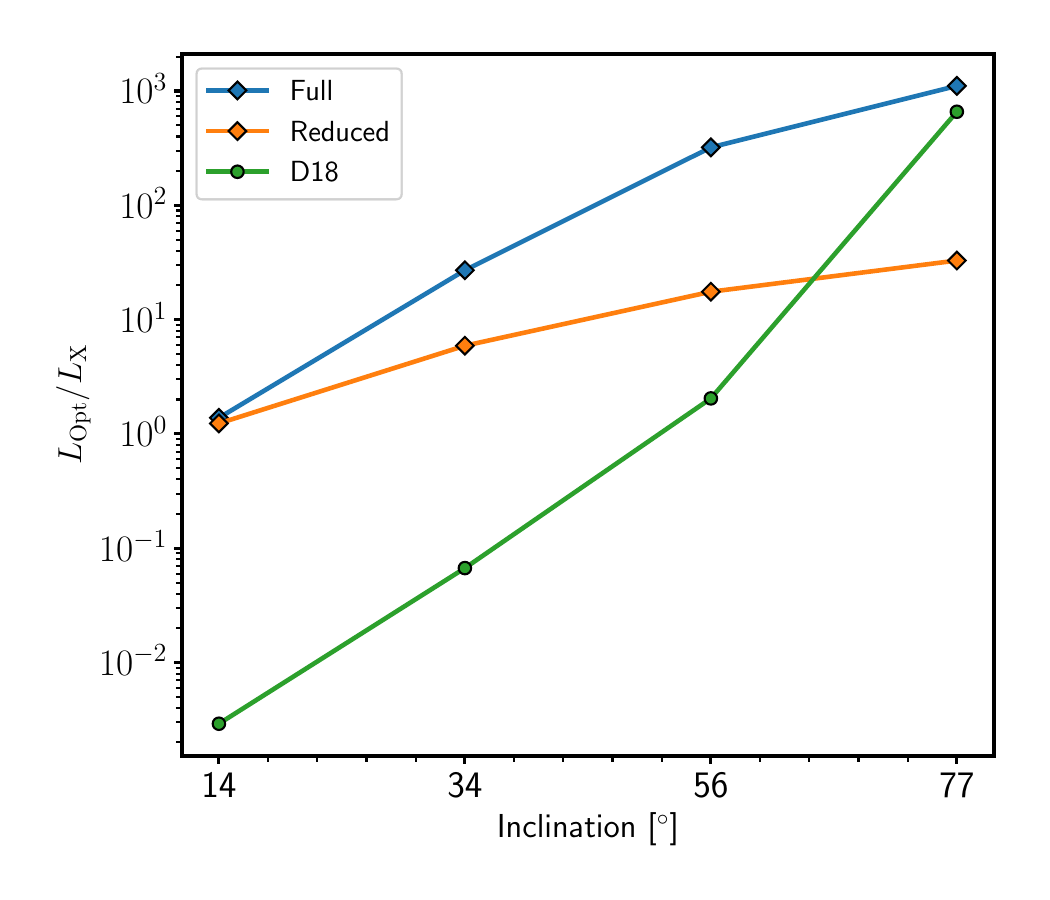}
    \caption{The integrated OXR luminosity ratios as a function of inclination angle for \citetalias{dai_unified_2018}'s quasi-1-D and our 2-D simulations. The X-ray band has been defined to include photons with energies 0.2 - 10 keV. The optical band is defined between $1700$ - $6500$ \AA. The OXR depends much more strongly on inclination in the quasi-1-D simulation, whereas the 2-D results are not as sensitive.} 
    \label{fig: oxr_multi_d}
\end{figure}

In the \citetalias{dai_unified_2018} unified model, the OXR ratio increases with inclination. This led to their proposal that X-ray bright TDEs correspond to systems observed at low (polar) inclinations, where X-ray emission escape along an optically thin funnel region. Optically bright TDEs, on the other hand, correspond to systems which are observed at high inclinations along the mid-plane, where the SED/X-ray emission has been reprocessed by the optically thick outflow resulting in an optically enhanced spectrum. This result is also found by \citetalias{Thomsen_2022}, who, expanding on the original results of \citetalias{dai_unified_2018}, conduct three new GRRMHD and MCRT post-processing simulations for different Eddington accretion ratios. They find that the inclination of an observer is the main parameter influencing the OXR ratio.

Figure \ref{fig: oxr_sed} shows synthetic spectra generated for five sight lines for our simulations, as well as two coloured regions representing the X-ray and optical bands. To quantify the inclination dependence of the OXR ratio, we have defined two X-ray bands. The first band includes photons with energies of 0.2 - 10 keV, and the second band includes photons with energies of 0.1 - 10 keV. We include two X-ray bands as the OXR is very sensitive to the band definition, as well as the frequency distribution of the input SED. We define the optical band as the wavelength region between $1700$ - $6500$ \AA.

Figure \ref{fig: oxr_sed} shows that the short wavelength emission decreases as inclination increases, with almost no emission coming from the X-ray band where $i \gtrsim 56$\degrees. Interesting to note is how the optical emission is fairly insensitive to inclination. This is an effect of multi-D photon transport, as reprocessed photons from the mid-plane of outflow are able to escape along sight lines of low optical depth. In the low inclination spectra, where $i \gtrsim 14$\degrees, there is increased X-ray emission especially when looking directly into the funnel region. This is from photons which scatter in the inflowing regions of the flow and are blueshifted, which then escape through the (jet evacuated) optically thin funnel region without being reprocessed. In Figure \ref{fig: oxr} we show the values of the OXR ratio as a function of inclination for two simulations (the ``reduced'' simulation will be discussed in more detail in Section \ref{sec: abund-comp}) and two X-ray band definitions. This shows, clearer, that the spectra become optically dominated as the inclination increases.  

Both Figure \ref{fig: oxr_sed} and Figure \ref{fig: oxr} suggest that our 2-D results are qualitatively consistent with the trend identified by \citetalias{dai_unified_2018}. However, there are some caveats. When we calculated the OXR using X-ray band 1 (0.2 - 10 keV) there is no inclination where the X-ray luminosity is significantly greater than the optical. We find a ratio of roughly unity at the lowest inclination in Figure \ref{fig: oxr}, where as \citetalias{dai_unified_2018}'s Bin 4 (corresponding to 14\degrees\ in our 2-D simulations) has an OXR of $0.003$ (see Figure \ref{fig: oxr_multi_d} below). However, there are inclinations when we use X-ray band 2 (0.1 - 10 keV) where the X-ray emission dominates with an OXR of $\sim 0.1$ for the 14\degrees\ measurement.

{In Figure \ref{fig: oxr_multi_d}, we directly compare the OXRs for both of our 2.5-D simulations with those obtained from \citetalias{dai_unified_2018}'s quasi-1-D calculations. Before discussing this comparison, it is important to note that the input SED in our 2.5-D simulations is characterized by different BB temperature ($T_{\mathrm{BB}} = 2\times10^{5}$ K; see Section~\ref{sec: radiation_sources}) than that in \citetalias{dai_unified_2018} ($T_{\mathrm{BB}} = 2\times10^{6}$ K). This difference affects the predicted OXRs, since these two input SEDs produce very different amounts of X-ray emission. In Section \ref{sec: multi_d_transport}, we directly compare our 2.5-D simulations to a set of quasi-1-D models that use the {\em same} input SED (i.e. $T_{\mathrm{BB}} = 2\times10^{5}$ K). The resulting spectra are shown in Figure~\ref{fig: multi_d_seds}, and these quasi-1-D models do not produce {\em any} X-ray emission.}

{Returning to Figure~\ref{fig: oxr_multi_d}, we see that there is a larger variation in the OXR -- and a much stronger dependence of the OXR on inclination -- in the 1-D simulations compared to the 2.5-D simulations. Even allowing for the differences in the input SEDs, 
this suggests that inclination alone may not be enough explain the wide range of variation in observed TDE OXRs \citep[see, e.g.,][]{saxton_correction_2021, Guolo_2024}. In particular, \citet{Guolo_2024} show that there is a variation in the OXRs at early times ranging from $\sim 10^{-2} - 10^{3}$, which is the epoch represented by \citetalias{dai_unified_2018}'s GRRMHD simulation, and consequently our post-processing too. This range of OXR is not seen in our simulations -- which only probes the inclination dependence of the OXR -- as the smallest OXR is $\sim 1$. There are no inclinations where X-ray emission dominates. The OXR range is also significantly smaller in the 2.5-D ``reduced'' simulation, where we have included only three atomic species.}

{Simulations by \citetalias{Thomsen_2022} have been used to propose a quasi-1-D \textit{dynamical} unification model. In their simulations, they found that whilst inclination is the primary parameter which influences the OXR, the OXR also correlates with the accretion rate, and therefore the mass-loss rate of the wind. In particular, they found that the OXR decreases as the accretion rate onto the central engine declines. Therefore TDEs can evolve from being optically to X-ray dominated, as the amount of obscuring/reprocessing material decreases. Our simulations support the idea that other variables are needed to explain the large dynamic range of observed OXRs.}

{It is important to keep in mind that we have only post-processed a single time-independent snapshot at a fixed Eddington ratio for a restricted range of inclinations. It could be that for this 2.5-D simulation we would only observe an X-ray dominated spectrum for extreme polar angles. Moreover, the variation of TDE OXRs also suggests that there could be a wide range of reprocessing rates in TDE outflows \citep{saxton_correction_2021}, which would not be captured in a single time-independent snapshot. Another important limitation is the use of single temperature blackbody to model the SED of the central engine. As noted above, this input SED will certainly impact the measured OXR. All that said, our results do confirm that multi-D photon transport impacts the OXR, since reprocessed photons can escape along paths of low optical depth. These photons then contribute to the UV/optical emission seen at low inclinations.}

\subsection{Reprocessing by the outflow} \label{sec: reprocessing}

\begin{figure*}
 \centering
 \includegraphics[scale=0.55]{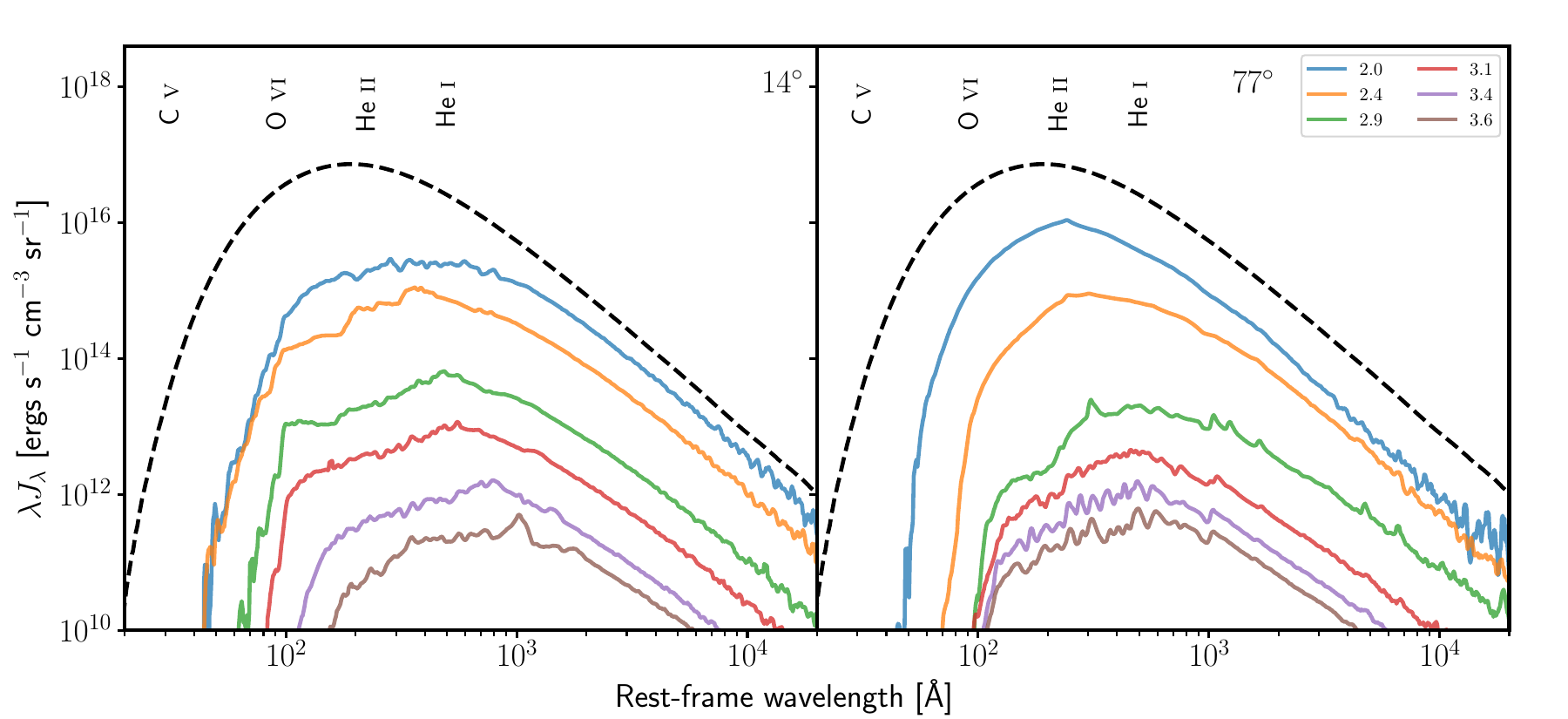}
 \caption{Rest-frame synthetic spectra generated for various grid cells at different logarithmic gravitational radii {(labelled with markers in Figure \ref{fig: model_properties})} along the two sight lines indicated in the top right corner of each panel. Each solid coloured SED represents the SED of a grid cell at that given radius and inclination. {The input SED is shown by the black dashed lined.} In general, the SEDs become weaker, broader and softer as the radius increases due to the radiation field becoming more dilute and reprocessed.}
 \label{fig: cell_seds}
\end{figure*}

 Reprocessing is a core component of the \citetalias{dai_unified_2018} unified model of TDEs, and for follow up simulations by \citetalias{Thomsen_2022}. It also is a core part of radiative transfer simulations carried out by \citet{roth_x-ray_2016} and \citet{parkinson_reprocessing_2022} which model the optical continuum and emission lines found in TDEs. Reprocessing is also important in our simulations, although it is more complex. 
 {Fundamentally, reprocessing via bulk scattering will tend to smoothly broaden and shift the SED, while atomic reprocessing will introduce more complex features (e.g. photo-ionization edges, line blanketing and individual strong emission/absorption lines).}
 
 To visualise how the SED is reprocessed, we show in Figure \ref{fig: cell_seds} the local SED along two sight lines, a characteristic low and high inclination, at multiple radial positions. The exact location of these cells are also indicated in Figure~\ref{fig: model_properties}. In general, the SEDs become weaker and softer as the radius increases, due to the radiation field becoming more and more dilute and reprocessed. {In interpreting these SEDs, it should be kept in mind that, in 2-D, the radiation field in any given cell can contain significant contributions from cells that do {\em not} lie along the line of sight to the central source.}

{The left panel of Figure \ref{fig: cell_seds} shows the evolution of the local SED along a low-inclination sight line. The SEDs of the cells located within $\log_{10}(r / r_g) < 3$ exhibit mainly atomic reprocessing. Due to the low densities in these regions, bulk scatter reprocessing is less important. {More specifically, the low electron scattering optical depth between the origin and these cells (relative to higher inclinations, see Figure~\ref{fig: optical_depth_surfaces}) means that photons emerging here will have typically undergone far fewer scattering events (in the optically thick limit, $N_{\rm scat} \propto \tau^{2}$) for both spherical and slab-like geometries \cite[e.g.][]{scatter}.)} On the other hand, the electron temperatures and ionization states in these cells are low enough for metals to retain some of their electrons, so the optical depths associated with several photo-ionization edges are sufficient for atomic reprocessing to play a role. {Figure \ref{fig: optical_depth_spectra} shows that even along these lower density sight lines, the continuum opacity for X-ray and EUV photons is large, with $\tau \gtrsim 100$. The cumulative effects of  bulk scatter reprocessing do become more apparent as we move towards larger radii along this sight line ($\log_{10}(r / r_g) > 3$), since here the velocities are largest (Figure \ref{fig:rebinned_profiles}). These cells do exhibit a broadening of the SED together with a clear shift to longer wavelengths.}}

{The evolution of the local SED along a high-inclination sight line is shown in the right panel of Figure \ref{fig: cell_seds}. {The SEDs of the cells located within $\log_{10}(r / r_g) < 3$ are affected little by atomic reprocessing, but are strongly affected by bulk scatter reprocessing. This leads to strong bulk scatter reprocessing, as is evident from the clear shift of these SEDs to the red, coupled with the absence of strong photo-ionization edges or emission/absorption lines characteristic of atomic reprocessing. This behaviour is caused by the rapid rise in the electron scattering optical depth along this sight line, with $\tau_{\rm{es}} \gtrsim 10$ at $\log_{10}(r / r_g) \sim 2.9$ and $\tau \gtrsim 100$ at $\log_{10}(r / r_g) \sim 2.4$. } There a strong change in the shape of the SED between $\log_{10}(r / r_g) = 2.9$  and $\log_{10}(r / r_g) = 3.1$. The latter is strongly attenuated relative to the former, with clear atomic reprocessing features (photo-ionization edges and emission lines). This transition in the SED shape is due to the sight line passing through a dense, cool and relatively low-ionization zone around $\log_{10}(r / r_g) \simeq 3$; {see Figure~\ref{fig: model_properties} and the continuum optical depths for this sight line in Figure \ref{fig: optical_depth_spectra}}. For example, this region has by far the highest \atomictransition{He}{ii} ionization fraction across the entire computational domain, which explains the appearance of the corresponding photo-ionization edge and absorption near 54~eV in the SED.}

 In broad terms, the reprocessing in our 2.5-D simulation is similar to the reprocessing in the quasi-1-D picture presented by \citetalias{dai_unified_2018} and \citetalias{Thomsen_2022}. Although, the non-spherical outflow structure and multi-D photon transport has made the reprocessing picture far more complicated. One such complication is how reprocessed photons are able to -- and do -- travel and escape along any sight line. Photons can do this as they are not forced to scatter through an optically thick barrier as they are in spherical geometries. Instead, they either scatter around the obstacle or travel in a completely opposite direction with lower opacity. Whilst there are \textit{still} distinct regions where different reprocessing mechanisms dominate, those reprocessed photons can be anywhere in the outflow and contribute to the local SED of each cell. This means that reprocessed photons in the quasi-1-D framework which could only contribute to the escaping radiation in one bin now contribute to each spectrum in our simulation. This is demonstrated by how the optical continuum is fairly insensitive to inclination; refer again to Figure \ref{fig: oxr_sed}. Optical photons generated by atomic reprocessing, for example, escape along the mid-plane and also polar inclinations, which is not possible in \citetalias{dai_unified_2018}'s 1-D setup.

\subsection{How important is multi-D radiative transfer?} \label{sec: multi_d_transport}

\begin{figure*}
\centering
 \includegraphics[scale=0.55]{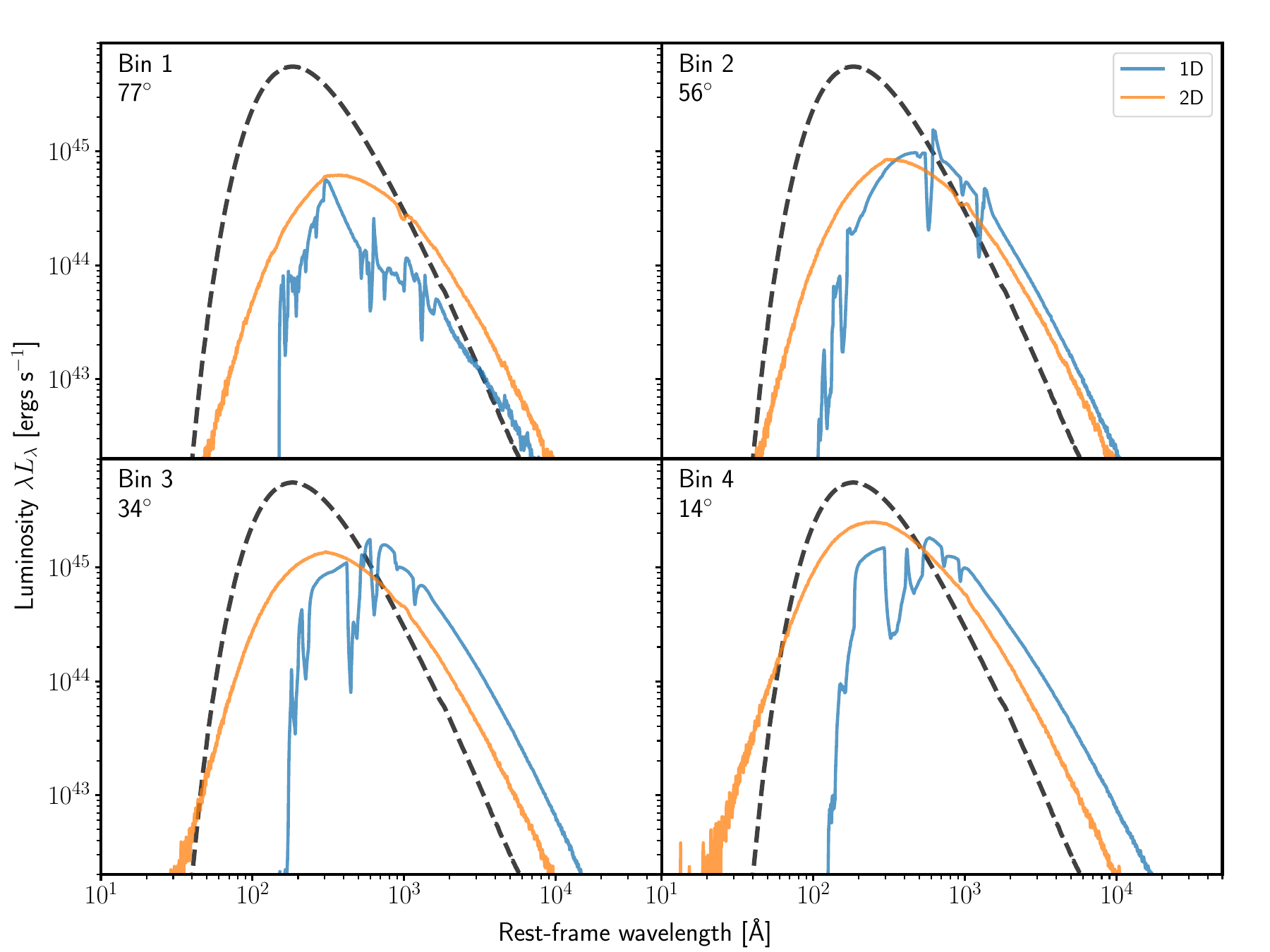}
 \caption{Rest-frame synthetic spectra of the 2.5-D simulation and a spectra of a comparable 1-D simulation using the same bins as \citetalias{dai_unified_2018}, but with a modified input SED and outer boundary. Both the 1-D and 2.5-D simulations include only H, He and O and have the same input SED. The 2.5-D spectra are picked to be the respective mid-point inclination angles for the opening angles of each 1-D simulation by \citetalias{dai_unified_2018}. Shown by a black dashed line is the input SED. The emergent spectra in 1-D and 2.5-D are broadly similar, appearing as reprocessed blackbodies which have been stretched and redshifted to longer wavelengths. However, the spectrum is reprocessed far more in 1-D due to the lower ionization state and lack of multi-D photon transport, which does not allow photons to escape along sight lines of lower optical depth and avoid repeated reprocessing.}
 \label{fig: multi_d_spectra}
\end{figure*}

\begin{figure*} 
 \centering
 \includegraphics[scale=0.55]{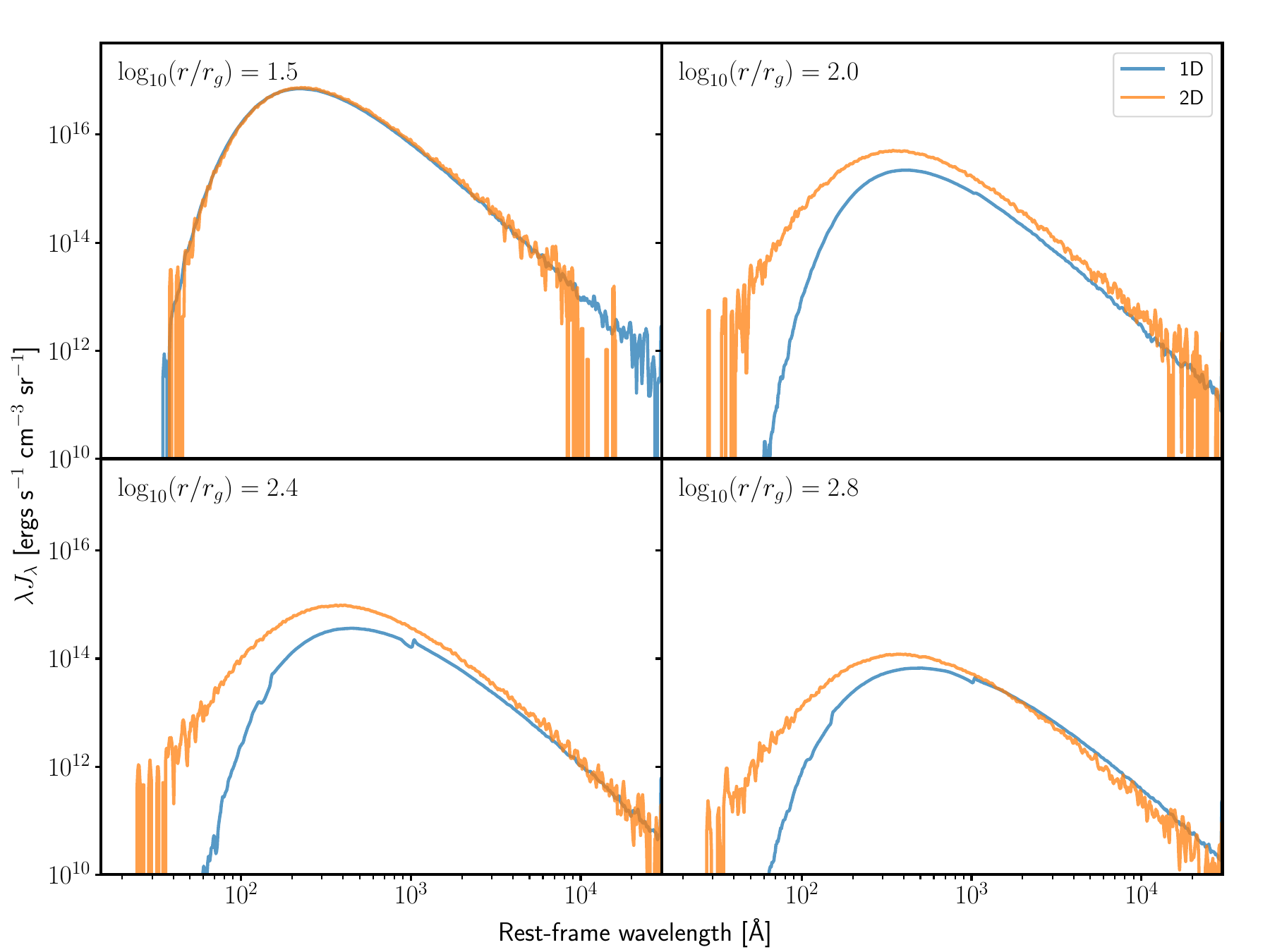}
 \caption{Rest-frame synthetic spectra generated for multiple grid cells for a 1-D and 2.5-D model along a 14\degrees sight line, corresponding to the spectra in the lower right panel of Figure \ref{fig: multi_d_spectra}. The radius of the grid cells is labelled in the top left of each panel. The input SED is reprocessed more in 1-D, evidenced by the atomic features in the 1-D spectra and short wavelength absorption which is not present in the 2-D results.}
 \label{fig: multi_d_seds}
\end{figure*}

 The purpose of this work is to extend previous 1-D post-processing simulations by \citetalias{dai_unified_2018} and \citetalias{Thomsen_2022} into 2.5-D, exploring how multi-D photon transport, in a non-spherical outflow, affects the inclination dependence of reprocessing by the outflow and the OXR. So, just how important is multi-D radiative transfer and how limiting is a spherical symmetry? {To explore this question, we simulate with \thecode\ the four $\theta$-binned \textit{spherical} simulations by \citetalias{dai_unified_2018}, with some modifications. We compare the emergent spectra from these spherical simulations with those generated from our 2.5-D simulation. To ensure a more meaningful comparison, we adjust the original setup by \citetalias{dai_unified_2018}, including modifications to the inner and outer boundaries and the input SED to align with the parameters of our 2.5-D simulation. \citetalias{dai_unified_2018} uses an input SED characterised by a blackbody of temperature of $T_{\mathrm{BB}} = 10^{6}$ K with a luminosity set such that the escaping luminosity is $\approx 2 L_{\mathrm{Edd}}$. In our 2.5-D simulation, the input SED is characterised by a blackbody of temperature $T_{\mathrm{BB}} = 2\times10^{5}$ K (which is the radiation temperature at our selected inner boundary of the GRRMHD simulation) with a luminosity $L_{\mathrm{BB}} \simeq 12~L_{\mathrm{Edd}}$.} In both setups, we include only H, He and O to simplify the comparison. We discuss the implications of different atomic abundances later in Section \ref{sec: abund-comp}. {In Appendix \ref{sec: python_verification}, we include a detailed description on how the computational grids for the spherical simulations are constructed. We also include a direct comparison between \thecode~and \citetalias{dai_unified_2018} using their original parameters; e.g. the original input SED and inner/outer boundaries.}
 
 {A comparison between the synthetic spectra obtained for the 1-D and 2-D simulations is shown in Figure \ref{fig: multi_d_spectra}, including the input SED. Note that our new 1-D simulations produce different spectra compared to the original \citetalias{dai_unified_2018} simulations. They are not as smooth and include line features not present in \citetalias{dai_unified_2018}. This is due to the changes of the inner/outer boundary and input SED. Most notably the larger outer radius adds in \textit{additional} (reprocessing) material not present in the \citetalias{dai_unified_2018} simulations. Ignoring the details of the spectra in Figure \ref{fig: multi_d_spectra}, there is a similar qualitative reprocessing picture between the two geometries. The spectra takes on the form of reprocessed blackbodies; discussed earlier in Section \ref{sec: synthetic_spectra}. The top left panel, showing the equatorial spectra, includes evidence of atomic reprocessing due the strong absorption of the SED and the absorption edges present. The bottom right panel, by contrast, represents a regime dominated by bulk scatter reprocessing, evident from the SED shift to longer wavelengths due to repeated scattering, with significantly fewer absorption and line features, particularly in 1-D.}

 This is where the similarity between the two geometries ends. In each panel of Figure \ref{fig: multi_d_spectra}, the details of the 1-D and 2-D spectra are very different, with more and stronger atomic features in 1-D. {These differences arise from the contrasting geometries and radiation fields, which influence the thermal and ionisation states of the outflows, as well as photon propagation and reprocessing, ultimately shaping the final converged states.} In all panels, the input SED is reprocessed much more in the spherical simulations. Not only are the spectra shifted further toward longer wavelengths via bulk scatter reprocessing, there are also strong absorption and emission lines as well as absorption edges which are not in the 2.5-D spectra. The 1-D spectra also do not have a high-energy tail as in the high inclination spectra in the 2.5-D simulation, as it has been absorbed via atomic reprocessing. 

 To show the difference in reprocessing between the geometries, in Figure \ref{fig: multi_d_seds} we show a comparison of the local SED, for four grid cells, for the Bin 4 simulation and the 14\degrees sight-line of the 2.5-D simulation (bottom right panel). This figure shows that SED is being modified \textit{more} by the spherical outflow. In each grid cell, the SED in the spherical simulations are weaker and peak at lower frequencies. They also exhibit edges from photoionization edges and the high-energy part of the SEDs are significantly more absorbed. There is also a stronger optical continuum in the furthest out spectra in the bottom right panel. 

 Whilst these differences are, in part, a symptom of the different ionization and thermal states, the fact that photons can easily become \textit{trapped} in spherical geometries when the scattering optical depth in moderate to high is also incredibly important\footnote{{Photon trapping also contributes to shaping the radiation field, which in turn influences the ionisation state.}}. In a non-spherical outflow, like our 2.5-D simulation, photons can -- and preferentially do -- travel around high optical depth, so photon trapping in much less efficient than in a non-spherical geometry. In spherical geometries, there is simply no way around a optically thick barrier barrier, so photons have to push through and  can become trapped. When a photon becomes trapped, it is reprocessed again and again and it interacts with the outflow. In our spherical simulations, the scattering optical depths are high causing photons to interact with the outflow multiple times. This means more bulk scatter reprocessing shifting the SEDs to longer wavelengths, and more absorption of high-energy photons by the dense regions. The spectrum for the spherical simulation in the top left panel of Figure \ref{fig: multi_d_spectra}, in particular, is most affected by photon trapping resulting in a heavily reprocessed spectrum. The spherical simulation has a lower scattering (and continuum) optical depth than the equivalent sight-line in the 2.5-D simulation, but because photons \textit{have} to push through the optical thick barrier, they scatter more and are more likely to be absorbed (and re-emitted at longer wavelengths).

 In the spherical simulations, the bolometric luminosity is roughly similar across all the synthetic spectra. But there is a dependence of the bolometric luminosity on inclination in the 2.5-D simulation, with the luminosity decreasing with inclination: the luminosity is brightest when looking at polar inclinations. This is another side effect of multi-D photon transport in a non-spherical outflow and is caused by photons preferentially travelling along sight lines of low optical depth which results in a beaming effect along polar directions; e.g. Figure \ref{fig: optical_depth_spectra} and Figure \ref{fig: optical_depth_surfaces} show there is less optical depth along low inclinations. 

 Overall, these two side-effects suggest that multi-D photon transport through a non-spherical outflow matters. Not only does it matter in terms of the \textit{detailed} reprocessing picture, it also introduces a new inclination dependence on the bolometric luminosity. Whilst quasi-spherical models can, and do, produce an inclination dependent unified model for optical and X-ray TDEs, it's clear from our results that detailed modelling in the future requires multi-D photon transport in non-spherical outflows.

\subsection{How important are the adopted abundances?} \label{sec: abund-comp}

 \begin{figure*}
     \centering
     \includegraphics[scale=0.55]{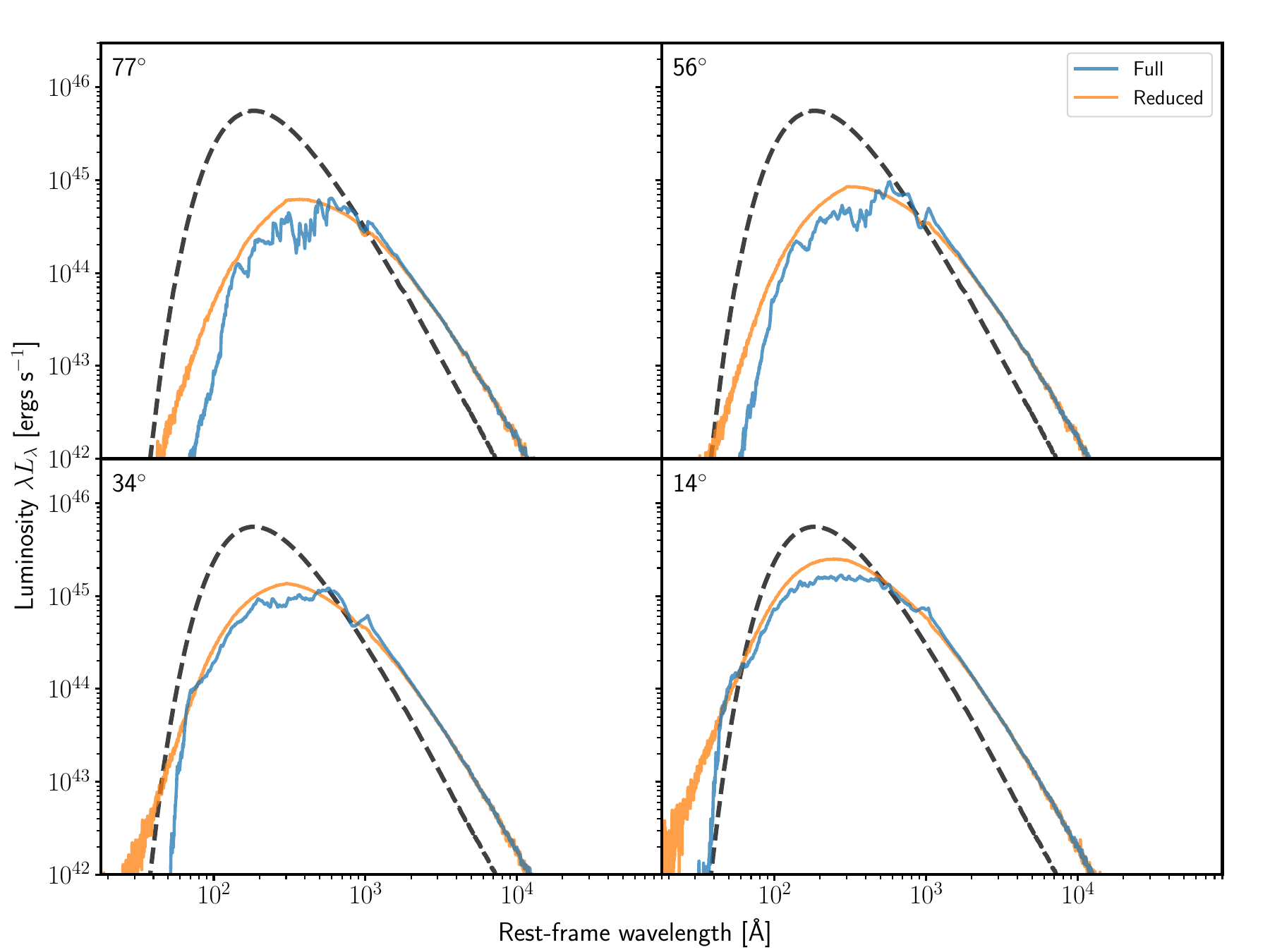}
     \caption{Rest-frame synthetic spectra for a simulation using the atomic data outlined in \ref{sec: atomic} and a reduced atomic data set containing only H, He and O. Shown by a black dashed line is the input SED. The broad qualitative reprocessing picture is the same between the atomic datasets. However, the inclusion of additional metal opacity results in numerous EUV emission and absorption lines.}
     \label{fig: model_spec_abun}
 \end{figure*}

 The radiative transfer by \citetalias{dai_unified_2018} considered the opacity and ionization from only H, He and O. In our simulation we have included more metals, increasing the number of routes for photons to be radiatively reprocessed. To explore how important the adopted atomic abundances are, we create synthetic spectra for a modified simulation where we include only H, He and O. We refer to this as the \textit{reduced} simulation, whereas the original is known as the \textit{full} simulation.

 Synthetic spectra of the two simulations are shown in Figure \ref{fig: model_spec_abun} for four sight lines. Broadly speaking, the reprocessing picture is the same between the atomic datasets. Both simulations produce qualitatively similar reprocessed SEDs, sensitive to the inclination of the observer and are therefore consistent with the \citetalias{dai_unified_2018} unified model -- this is also shown in Figure \ref{fig: oxr}. But there are a number of detailed differences to consider. The first is that the inclusion of more metals results in a family of EUV emission and absorption lines, which are missing in the reduced simulation. The additional metal opacity in the full simulation also results in increased absorption and reprocessing of high energy photons. Figure \ref{fig: oxr} also shows this, with the full simulation producing orders of magnitude larger OXRs for the 0.2 - 10 keV band for $i > 14$\degrees, resulting in a larger dynamic range of OXRs. In the low inclination spectra for both simulations, there are high energy tails {produced by scattering with high velocity electrons.} This is absorbed in the full simulation, but it is not the case in the reduced abundance simulation. {At wavelengths longer than $1000$ \AA, the inclusion of the additional metals in the full abundance simulation has little impact on the spectra. This is because the metals do not provide significant (continuum) opacity to photons in this wavelength range (refer back to Figure \ref{fig: optical_depth_spectra}).}

 Future detailed modelling to compare theory to observation therefore requires, according to the results here, as detailed as possible ionization and radiative transfer, which also means using realistic atomic abundances. Differences to the ionization structure, opacity and from photon transport all contribute to the detailed structure of the synthetic spectra generated. {Given that we have included all of the dominant atomic species, we do not expect the spectra to change by much with more metals.} Since the shape of the SEDs between the two setups is broadly similar, the reduced simulation -- which is easier to interpret -- is still useful for the purposes of developing physical intuition.

\subsection{Limitations} \label{sec: limitations}

 Whilst we have been able to relax some approximations and assumptions made in previous post-processing simulations, there are still several which remain. The main approximation we have not been able to relax is the radiation source. The input radiation field is modelled by including an isotropically emitting central source, with an SED governed by a single temperature blackbody distribution. If the emission comes from an accretion disk, then a single temperature blackbody is a poor approximation for the shape of the SED and the angular distribution is likely to be rather anisotropic. Whilst our choice of radiation source is a reasonable first approximation, a more realistic model of the continuum is required to obtain quantitative, rather than qualitative, results to compare with observations. However, the structure, properties and emissions of radiation dominated and vertically extended accretion disks are still an area of on-going research \citep[e.g.][]{hirose_radiation-dominated_2009,jiang_thermal_2013, Shen2014}. 

 We have assumed that the outflow is in radiative near-equilibrium and in a steady state. However, in a handful of TDEs, such as AT2018zr \citep{hung_discovery_2019} and AT2019qiz \citep{nicholl_outflow_2020, hung_discovery_2020}, the optical and UV spectra have been observed at early times to change over timescales as short as 10-20 days. The snapshot we have post-processed represents the $\Delta t \approx 5$ days epoch, when inflow equilibrium of the disc and only a quasi-steady state has been reached. By post-processing this time-averaged snapshot, the evolution of the outflow, including its effect on the emergent light, is completely ignored. An obvious next step would be to post-process multiple snapshots at different time snapshots/evolutionary stages, in a vein similar to \citetalias{Thomsen_2022}. The outflow is also still limited to two dimensions (without rotational velocity). Whilst this is an advancement on the earlier 1-D treatments, it is still an incomplete picture.

 Finally, we have neglected general relativistic effects. In \thecode, there is no concept of an event horizon; although it is crudely approximated by having an absorbing inner boundary. Photons also travel in straight lines between interactions. Close to the SMBH, the curved paths of photons could subtly change the value of Monte Carlo estimators (such as the mean radiation field or heating and cooling rates) from which the properties of the outflow and its ionization state are determined by. However, the majority of reprocessing happens at $r / r_{g} \approx 500$, where GR effects are less likely to matter, so this is perhaps a minor concern compared to other approximations in our work. 

\section{Summary} \label{sec: summary}

 The reprocessing of disc emission by an optically thick outflow, or extended envelope, has been a promising mechanism to explain the prevalence of optically bright, but X-ray weak, TDEs at early times. It is, in-fact, a core feature of the unified model suggested by \citetalias{dai_unified_2018} which is based on quasi-1-D framework of post-processed MCRT simulations, constructed from their 3-D GRRMHD TDE simulation. The primary aim of our work was to explore to what extent post-processing in a 2.5-D framework impacts the conclusions obtained from the same GRRMHD snapshot. As in \citetalias{dai_unified_2018}, we find that reprocessing by a (2-D) outflow can explain the existence of both X-ray bright and optically bright TDEs, with inclination as a critical parameter. That said, the inherently multi-D nature of the outflow significantly affects the radiative transfer, modifying the emergent spectra and the inclination dependence of the bolometric luminosity and the OXR. Our main results are:
 
 \begin{enumerate}
    \item The orientation-dependent OXR emission ratio increases with inclination angle in our 2.5-D simulations, consistent with \citetalias{dai_unified_2018}'s unified model. However, we do not find any inclinations which are significantly brighter in the X-ray. The OXRs also do not depend as strongly on the inclination as in \citetalias{dai_unified_2018}'s simulations.

    \item Our 2.5-D simulations also show a significantly smaller dynamic range of OXRs compared to observed TDE OXRs. We suggest that inclination alone may not account for the observed variation and additional variables may be needed to explain the full dynamic range of observed OXRs.

    \item Whilst the qualitative reprocessing picture remains similar to \citetalias{dai_unified_2018}'s quasi-1-D framework, multi-D radiative transfer through a non-spherical geometry modifies the details of how emission is reprocessed. In a non-spherical outflow, photon trapping and reprocessing are far less efficient. Photons can, and do, preferentially escape along paths of lower optical depth. This suggests that future modelling needs to account for multi-D velocity and ionization structures, as well as multi-D photon transport.

    \item While electron scattering dominates the overall opacity ($\tau \sim 10-100$) throughout most of the flow, photoionization processes are the dominant opacity source in the equatorial region of the outflow, where it is most dense and least ionized.

    \item Repeated electron scattering broadens the spectral energy distribution and shifts it redwards towards optical wavelengths, producing an emergent spectrum reminiscent of a stretched blackbody. Photons absorbed in the dense base, due to photo-absorption, are re-emitted at longer (UV/optical) wavelengths via recombination. 

    \item The weaker dependence of the OXR on inclination is mainly caused by multi-D photon transport. Photons which are reprocessed in the equatorial region can escape along polar directions of lower optical depth, something which is impossible in a spherical outflow. Therefore optical photons, as well as bare disc emission, reaches polar observers.
    
 \end{enumerate}

 To summarise, the 2.5-D radiative transfer simulations we have carried out with \thecode\ produce synthetic spectra which are qualitatively consistent with those predicted by the unified model proposed by \citetalias{dai_unified_2018}. They support the idea that reprocessing of disc emission, by an optically thick outflow, can produce the varied X-ray and optical emission properties observed in TDEs. However, the spectra and measured OXRs from our 2.5-D simulation are substantially different to those from \citetalias{dai_unified_2018}'s quasi-1-D simulations, suggesting that multi-D radiative transfer will be an important parameter for future studies and quantitative modelling of TDEs.
 
\section*{Data availability}

 The materials required to reproduce the simulations and the figures in this paper (including simulation outputs) are publicly available online at \href{https://github.com/saultyevil/tde_multi-D_unified}{\texttt{https://github.com/saultyevil/tde\_multi-D\_unified}}.

\section*{Acknowledgements}

 We thank the anonymous reviewer for their careful reading of the manuscript, and their helpful and insightful feedback which significantly improved the quality of this work. Figures were prepared using \texttt{matplotlib} \citep{hunter_matplotlib_2007}. The authors acknowledge the use of the \texttt{GNU Science Library} \citep{galassi_gnu_2006}. The authors acknowledge the use of the IRIDIS High Performance Computing Facility, and associated support services at the University of Southampton. The authors acknowledge that \thecode\ was developed with assistance of Research Software Engineers from the Southampton Research Software Group, at the University of Southampton. EJP acknowledges financial support from the EPSRC Centre for Doctoral Training in Next Generation Computational Modelling grant EP/L015382/1. LD and LLT acknowledge the support from the National Natural Science Foundation of China and the Hong Kong Research Grants Council (12122309,  27305119, 17304821, N\_HKU782/23). Partial support for KSL's effort on the project was provided by NASA through grant numbers HST-GO-16489 and HST-GO-16659 and from the Space Telescope Science Institute, which is operated by AURA, Inc., under NASA contract NAS 5-26555.


\bibliographystyle{styles/mnras}
\bibliography{bibliography}


\clearpage
\appendix

\section{A Benchmark Test Case} \label{sec: python_verification}

\subsection{Setup}

\begin{figure}
\centering
\includegraphics[scale=0.39]{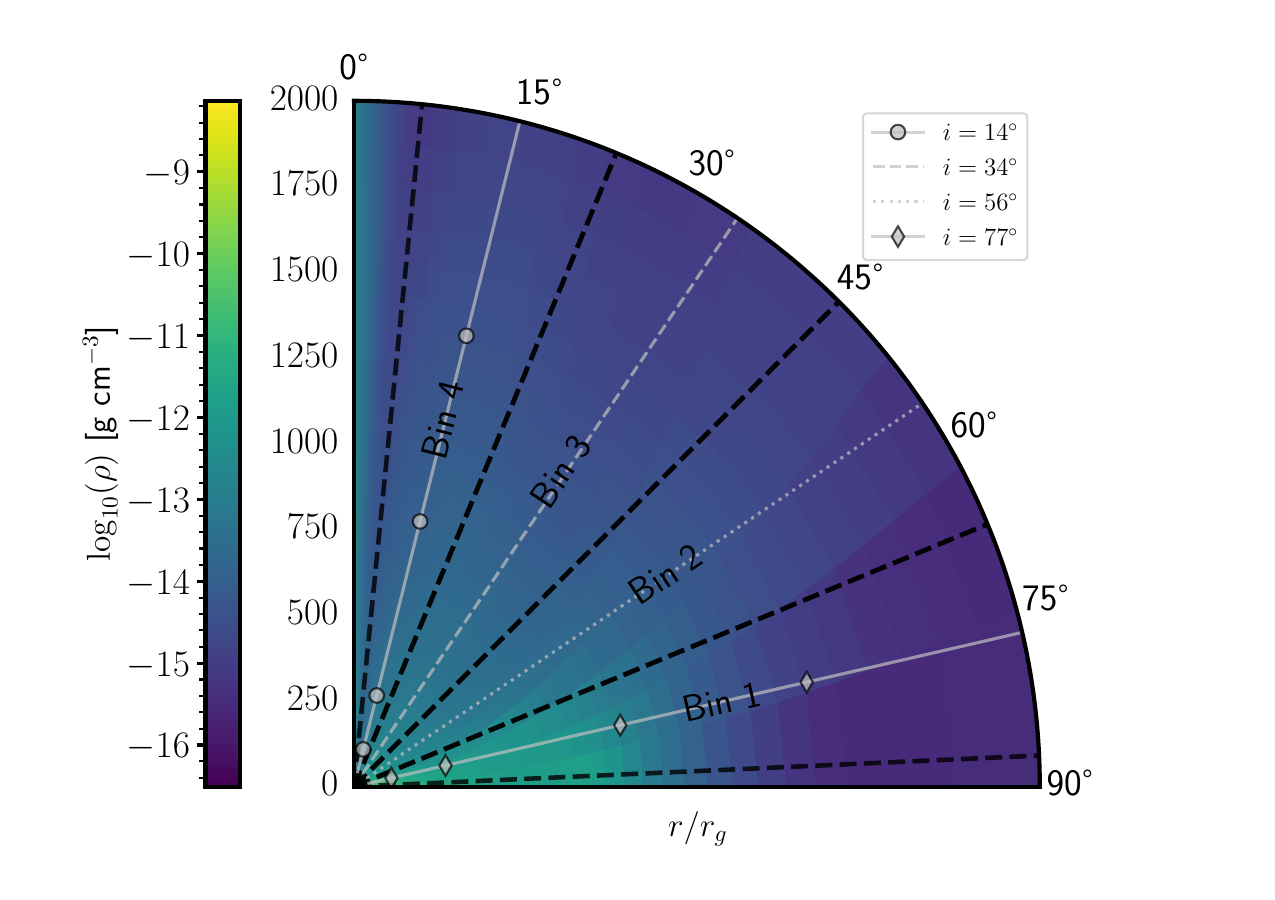}
\caption{A colour plot showing the mass density of the model from \citetalias{dai_unified_2018} for the region covering $0^{\circ} < \theta < 90^{\circ}$, where the mid-plane is located at $\theta = 90^{\circ}$. The regions labelled and bounded by the dashed lines are the four $\theta$-averaged models described by \citetalias{dai_unified_2018} in Section 2.2. {The grey lines, labelled in the legend, correspond to sight lines used in our 2-D simulation for each bin. The markers for the $i=14$\degrees~and $i=77$\degrees~sight lines correspond to the radial sample points in Figure \ref{fig: cell_seds}.}}
\label{fig: spherical_bins}
\end{figure}

To benchmark \thecode\  and our method against \citetalias{dai_unified_2018}, we have generated the same $\theta$-averaged simulations following the method outlined in \citetalias{dai_unified_2018} who used the MCRT code \sedona\ \citep{kasen_time-dependent_2006} to generate their spectra. We have created our 1-D simulations from our re-gridded grid described in Section \ref{sec: model_setup}, where Figure \ref{fig: spherical_bins} shows the region of the outflow occupied by each $\theta$-averaged 1-D grid. To transform each wedge into a spherical grid, the density and velocity are squashed into a single point by taking a volume weighted average in the $\theta$ direction. We set the inner boundary of each grid to the radial point where inflow equilibrium has broadly been reached, i.e. where the flow has transitioned from inflow to outflow. In some grid, especially near the mid-plane, the velocity field is complex and some cells are still in-flowing past the inner boundary. These cells do not contribute to the volume weighted averages.
 
\begin{figure}
\centering
\includegraphics[scale=0.4]{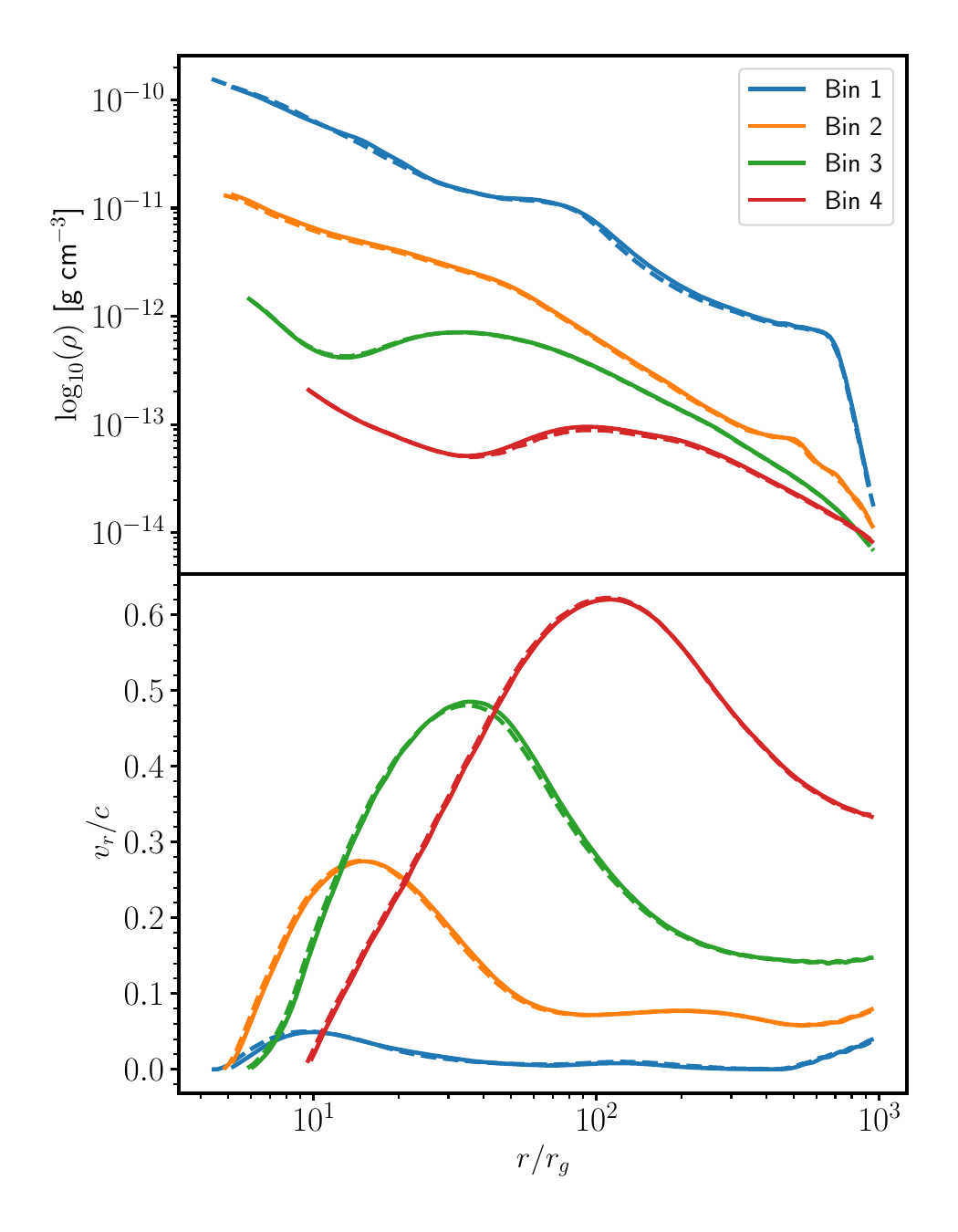}
\caption{Density and velocity profiles, as a function of $r$, for the $\theta$-averaged models in this work (solid) and those generated by \citetalias{dai_unified_2018}. The profiles shown are in excellent agreement. The small discrepancies are due to differences from remapping the warped unstructured grid onto a structured grid.}
\label{fig: spherical_properties}
\end{figure}

The outer boundary of each grid is set to a fixed value of $r / r_g = 1000$. This cutoff was chosen in \citetalias{dai_unified_2018} because there is material beyond this radius in Bin 1 which was used to create an artificial reservoir to fuel the accretion processes. Including this material could artificially affect the reprocessing of photons. For the sake of consistency, this outer boundary is imposed for all grids. A comparison to the grids generated for this work and those used in \citetalias{dai_unified_2018} is shown in Figure \ref{fig: spherical_properties}, which shows excellent agreement.

Photon packets are injected from the inner boundary, which also acts as an absorbing inner boundary to remove photons and acts as a crude approximation for an event horizon; \thecode\  does not account for any GR effects. The distribution of photons is sampled from a $T_{\rm BB} = 10^{6}$ K blackbody, inspired by the radiation temperature at the inner boundary of the GRRMHD simulation. As in \citetalias{dai_unified_2018}, the luminosity of the central source is set to normalise the escaping luminosity to $\approx 2 L_{\rm{Edd}}$ escaping from from the Bin 1 and Bin 2 simulation. We also only include H, He and O for the ionisation calculations. The key parameters of our simulations are summarised in Table \ref{tab: theta_avg_models}.

\begin{table}
\centering
\begin{tabular}{ccccc}
\hline
\multirow{2}{*}{Model} & $\theta_{1}$ & $\theta_{2}$ & $r_{\rm{in}}$  & $L_{\rm{in}}$ \\
               & ($^{\circ}$) & ($^{\circ}$) & ($r_{\rm{g}}$) & ($L_{\rm{Edd}}$)      \\ \hline
Bin 1                  & 67           & 87           & 4.36           & 100           \\
Bin 2                  & 45           & 67           & 4.81           & 48            \\
Bin 3                  & 22           & 45           & 5.84           & 48            \\
Bin 4                  & 5            & 22            & 9.49           & 48           \\ \hline
\end{tabular}
\caption{Parameters describing the binning of the theta averaged simulations from \citetalias{dai_unified_2018}. Included in the table are the opening and closing angles $\theta_1$ and $\theta_2$, as well as the inner boundaries in gravitational radii ($r_{g} \approx 7.38 \times 10^{11}$ cm) and the luminosity of the radiation source in units of Eddington luminosity ($L_{\mathrm{Edd}} = 6.26 \times 10^{44}$ ergs s$^{-1}$).}
\label{tab: theta_avg_models}
\end{table}

\subsection{Results}

\begin{figure*}
\centering
\includegraphics[scale=0.55]{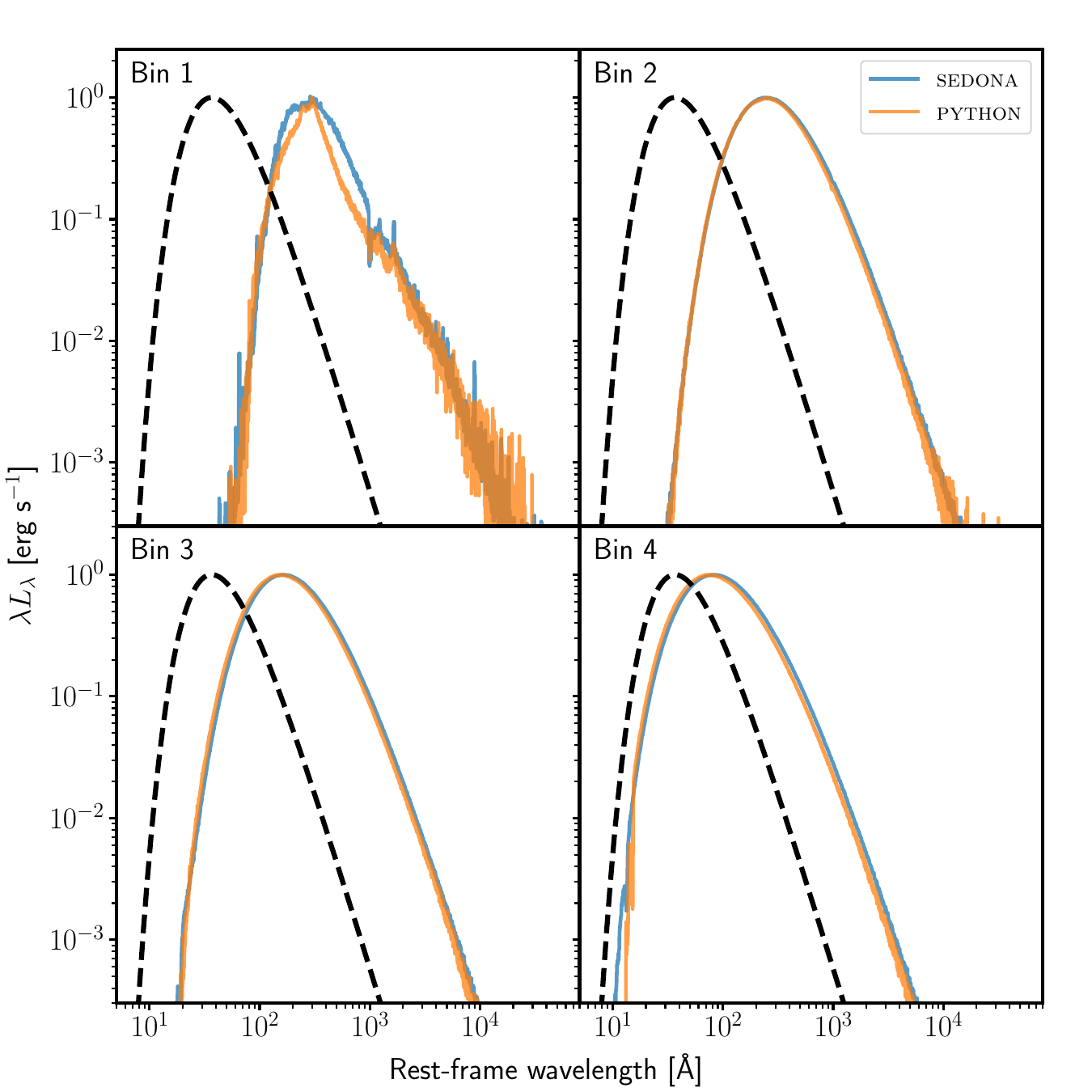}
\caption{Peak-normalised synthetic rest-frame spectra generated using a reduced atomic data set for the four 1-D $\theta$-averaged simulations. The same spectra presented by \citetalias{dai_unified_2018} created using \sedona are also included. Additionally, shown as a black dashed line is the input SED. The name of each simulation is labelled in the top left. There is generally excellent agreement between \thecode\  and \sedona, with the main differences in the Bin 1 simulation being due to differences in how atomic and thermal processes are modelled.}
\label{fig: spherical_spectra}
\end{figure*}

The synthetic spectra generated by \thecode\  are shown in Figure \ref{fig: spherical_spectra} alongside the spectra generated by \citetalias{dai_unified_2018}. For Bin 2, 3 and 4 we find excellent agreement between \thecode\  and \sedona. The emergent spectra take the shape of a stretched blackbody, which has been redshifted to longer, optical, wavelengths relative to the input SED. The redshifting, as in \citetalias{dai_unified_2018}, is caused by bulk scatter reprocessing, which reduces the mean photon energy of the photon population through successive scattering \citep{titarchuk_downscattering_2005, laurent_effects_2007, roth_what_2018}. Since H and He are almost completely ionized throughout the \thecode\  outflow, and O exists mostly in \atomictransition{O}{vii} and above (but is completely ionized for $r / r_{g} < 100$), there is very little absorption of the radiation. 

The agreement between Bin 1 is not as excellent, however it is still very good. The difference in this simulation is caused by differences in how \thecode\  and \sedona\ model atomic processes and balance heating and cooling processes, which has resulted in different thermal and ionisation states. This simulation is far more sensitive to differences in atomic physics because of the high densities and lower ionization state. Perhaps the most important numerical difference -- which is likely to be responsible for at least some of the differences in the predicted spectra -- is that \sedona\ does not include Compton heating and cooling in the thermal balance \citep{roth_monte_2015}. At least in Bin 1, Compton heating and cooling is important to the thermal balance in \thecode, which affects the ionization state of the outflow.


\bsp
\label{lastpage}
\end{document}